**Hydrocode modeling of the spallation process during hypervelocity impacts: Implications for the ejection of Martian meteorites**

Kosuke Kurosawa[a,*], Takaya Okamoto[a], and Hidenori Genda[b]

[a]Planetary Exploration Research Center, Chiba Institute of Technology, 2-17-1, Tsudanuma, Narashino, Chiba 275-0016, Japan

[b]Earth–Life Science Institute, Tokyo Institute of Technology, 2-12-1 Ookayama, Meguro-ku, Tokyo 152-8550, Japan

*Corresponding author

Kosuke Kurosawa

Researcher, Planetary Exploration Research Center, Chiba Institute of Technology

Tel: +81-47-478-0320

E-mail: kosuke.kurosawa@perc.it-chiba.ac.jp

**Abstract**

Hypervelocity ejection of material by impact spallation is considered a plausible mechanism for material exchange between two planetary bodies. We have modeled the spallation process during vertical impacts over a range of impact velocities from 6 to 21 km/s using both grid- and particle-based hydrocode models. The Tillotson equations of state, which are able to treat the nonlinear dependence of density on pressure and thermal pressure in the strongly shocked matter, were used to study the hydrodynamic–thermodynamic response after impacts. The effects of material strength and gravitational acceleration were not considered. A two-dimensional time-dependent pressure field within a 1.5-fold projectile radius from the impact point was investigated in cylindrical coordinates to address the generation of spalled material. A resolution test was also performed to reject ejected materials with peak pressures that were too low due to artificial viscosity. The relationship between ejection velocity $v_{eject}$ and peak pressure $P_{peak}$ was also derived. Our approach shows that "late-stage acceleration" in an ejecta curtain occurs due to the compressible nature of the ejecta, resulting in an ejection velocity that can be higher than the ideal maximum of the resultant particle velocity after passage of a shock wave. We also calculate the ejecta mass that can escape from a planet like Mars (i.e., $v_{eject} > 5$ km/s) that matches the petrographic constraints from Martian meteorites, and which occurs when $P_{peak} = 30$–$50$ GPa. Although the mass of such ejecta is limited to 0.1–1 wt% of the projectile mass in vertical impacts, this is sufficient for spallation to have been a plausible mechanism for the ejection of Martian meteorites. Finally, we propose that impact spallation is a plausible mechanism for the generation of tektites.

## 1. Introduction

Ejection of materials is an inevitable outcome of hypervelocity impacts onto planetary bodies, because such impacts induce a compression (shock) wave and a subsequent release (rarefaction) wave in the impactor and target planetary body. These waves accelerate materials and drive excavation flow [e.g., Melosh, 1985b]. The ejection process can be classified into three stages depending on ejection timing, location, velocity, and pressure: (1) jetting [e.g., Kieffer, 1977; Melosh and Sonett, 1986; Ang, 1990; Vickery, 1993; Sugita and Schultz, 1999; Johnson et al., 2014, 2015; Kurosawa, et al., 2015]; (2) spallation [Melosh, 1984, 1985a; Vickery and Melosh, 1987; Polanskey and Ahrens, 1990; Head et al., 2002; DeCarli et al., 2007]; and (3) normal excavation [e.g., Maxwell, 1977; Croft, 1980; Housen et al., 1983; Melosh, 1985b; Yamamoto and Nakamura, 1997; Cintala et al., 1999; Yamamoto, 2002; Anderson et al., 2004; Yamamoto et al., 2005; Hermalyn and Schultz, 2010; Housen and Holsapple, 2011; Johnson et al., 2013; Tsujido et al., 2015]. The ejection process gradually changes with time from jetting to spallation to normal excavation [e.g., Johnson et al., 2014; Kurosawa et al., 2015; Okamoto et al., 2016]. The ejection velocity and peak pressure experienced are different for these ejection processes depending on the geometric configuration during a shock-release sequence.

Jetting occurs at the earliest stage of excavation. Oblique convergence between a spherical projectile and target leads to a local energy concentration, resulting in an ejection velocity that is higher than the impact velocity. The mass ejected by jetting is estimated to be 0.1–1 wt% of the projectile mass [Johnson et al., 2014], which is relatively small compared with the other ejection mechanisms. The jetting process of spherical projectiles has been well studied by experimental, analytical, and numerical approaches [Melosh and Sonett, 1986; Vickery, 1993; Sugita and Schultz, 1999; Johnson et al., 2014; Kurosawa et al., 2015]. Normal excavation is driven by "the residual particle velocity" $u_{p\_res}$ as a consequence of shock release, which is produced due to irreversible shock heating [Melosh, 1985b]. The magnitude of $u_{p\_res}$ can be calculated by subtracting the velocity change due to pressure release $u_{p\_release}$ from the particle velocity under the shocked state $u_{pH}$. The procedures used to calculate $u_{pH}$ and $u_{\text{p\_release}}$ are described later in Section 2. Although the ratio of $u_{\text{p\_res}}$ to $u_{\text{pH}}$ depends on the accuracy of the equations of state (EOS) and the peak pressure experienced, in general $u_{\text{p\_res}}$ is much smaller than $u_{\text{pH}}$. Given that $u_{\text{pH}}$ is less than half of the impact velocity $v_{\text{imp}}$ in the case of collisions between two identical bodies, the ejection velocity produced by normal excavation is much smaller than the impact velocity. The main feature of normal excavation is that the ejected mass accounts for ~90 wt% of the whole ejecta during impact events. The relationships between the ejected mass and the launch position, velocity, and angle have been extensively investigated by experimental, analytical, and numerical approaches [e.g., Maxwell, 1977; Croft, 1980; Housen et al., 1983; Melosh, 1985b; Yamamoto and Nakamura, 1997; Cintala et al., 1999; Yamamoto, 2002; Anderson et al., 2004; Yamamoto et al., 2005; Hermalyn and Schultz, 2010; Housen and Holsapple, 2011; Johnson et al., 2013; Tsujido et al., 2015]. In contrast, ejection behavior due to spallation remains poorly understood [e.g., DeCarli et al., 2007; Melosh and Ong, 2011; Ong and Melosh, 2012; DeCarli, 2013]. In this study, we focus on the spallation process.

The spallation process has been investigated as a possible launch mechanism for lunar and Martian meteorites [e.g., Melosh, 1984, 1985a]. The most important feature of the spallation process is that lightly shocked ejecta fragments can be launched at relatively high velocities. This feature is consistent with petrographic evidence from Martian meteorites, which experienced relatively low peak pressures of 30–50 GPa [e.g., Head, 2002]. In addition, Martian meteorites must escape from Mars, and to do this their ejection velocity must exceed 5 km/s. Hereafter, conditions with peak pressures $P_{\text{peak}}$ = 30–50 GPa and ejection velocities $v_{\text{eject}}$ > 5 km/s are referred to as the Martian meteorite (MM) condition.

A lightly shocked, high-speed component was reported in a numerical computation for the first time by Ahrens and O'Keefe (1978). Melosh (1984) developed the first analytical model for impact spallation that describes the origin of such a component. Subsequently, Head et al. (2002) numerically modeled vertical impacts on a flat surface with the equations of state for geological materials, in order to investigate the material ejection due to spallation, and showed that the ejected materials meet the MM conditions. However, DeCarli et al. (2007) argued that the low peak pressures observed in the numerical calculations were computational artifacts that resulted from an artificial viscosity, which is necessary for capturing shock waves in hydrocodes. The artificial viscosity smooths a shock front over 3–10 computational cells [e.g., DeCarli et al., 2007; Johnson et al., 2014], resulting in an underestimated peak pressure near the free surface that may be 10%–30% of the true pressure of the shocked state, depending on the viscosity parameters [DeCarli, 2013]. Consequently, the impact outcomes from spallation, such as the mass–velocity–pressure distributions, are not fully understood. The three ejection mechanisms discussed throughout this section are illustrated in a schematic cross-section in Fig. 1. The jetting and spallation processes occur near the impact point, where the point source approximation is not valid. The near-surface wave interaction around the impact point is key to understanding the nature of the impact spallation process.

In this study, we have modeled vertical impacts using the 2-D iSALE shock physics code [e.g., Amsden et al., 1980; Ivanov et al., 1997; Collins et al., 2004; Wünnemann et al., 2006] and a 3-D Smoothed-Particle-Hydrodynamics (SPH) code [Lucy, 1977; Monaghan, 1992]. To explore whether spallation is able to launch lightly shocked materials from Mars, we carried out high-resolution simulations with up to 2000 CPPR for iSALE and 200 CPPR for SPH, where CPPR is the number of cells per projectile radius. Grid-based (iSALE) and particle-based (SPH) hydrocodes were used in this study to examine the inter-code variability on the impact outcomes. The effects of shock smearing near the free surface on ejection behavior were carefully investigated by conducting numerical calculations at different spatial resolutions. Thus, we were able to assess the significance of these artifacts by testing the resolution effects on the velocity–pressure relationship. We then derived the mass–velocity–pressure relationships for the spalled ejecta near the impact point.

The remainder of this paper is organized as follows. Section 2 describes difficulties in accounting for the launch of Martian meteorites during hypervelocity impacts based on shock physics, in order to clarify the problem to be solved by the hydrocode modeling. Section 3 presents the details of the numerical calculations and the results

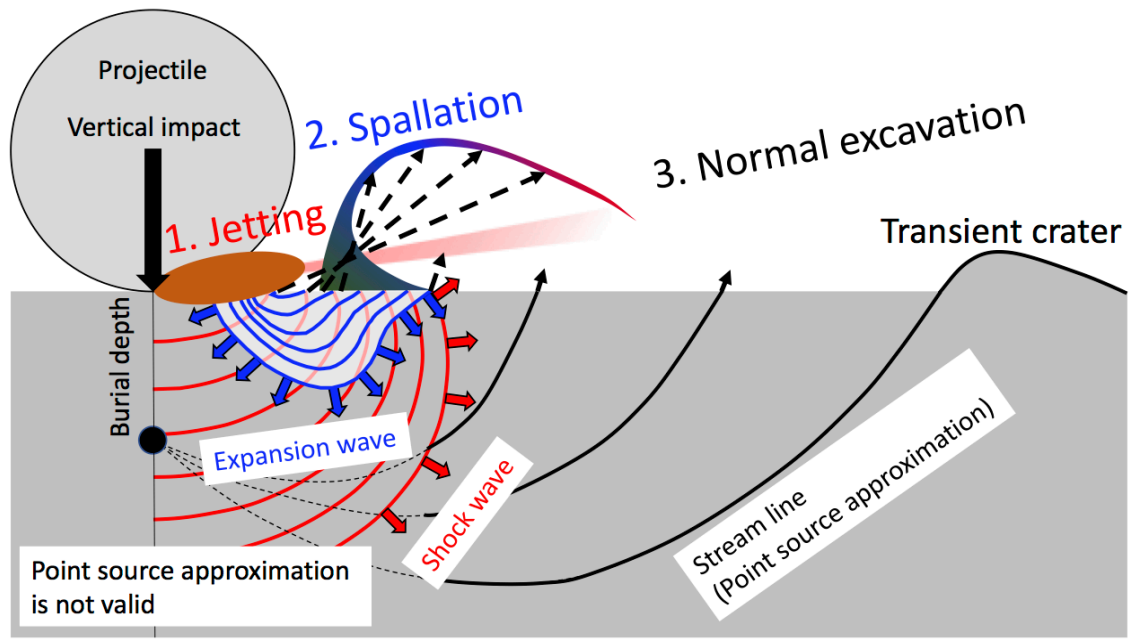

**Figure 1.** Schematic cross-section showing the transition of the ejection process from jetting to spallation and normal excavation. The isochrones for shock and expansion waves are from our numerical results (see Fig. 8a). The curving shape of the ejecta curtain is based on the curtain generated in impact experiments by Kurosawa et al. (2015). The curving shape originates from a gradual change in ejection angle during the transition from jetting to spallation. The cratering flow field due to normal excavation is based on Maxwell's Z model with an effective center of flow at a given burial depth, which is often referred to as the Z-EDOZ model [e.g., Croft, 1980; Kurosawa, 2015]

are presented in Section 4. In Section 5, we describe the nature of the spallation process revealed by our numerical simulations, the limitations of our model, and geological implications. Finally, our conclusions are presented in Section 6.

## 2. Difficulties in accounting for the launch of Martian meteorites

In this section, we discuss the role of thermal pressure in strongly shocked matter on material ejection, and the importance of numerical approaches to understanding the launch mechanisms of Martian meteorites. We first consider the analytical model for impact spallation proposed by Melosh (1984). This spallation model assumes a transverse stress pulse propagates from a burial depth $d$ with a finite rise and decay time, which are the same order of magnitude of the characteristic time for projectile penetration $t_s = D_p/v_{imp}$, where $D_p$ and $v_{imp}$ are projectile diameter and impact velocity, respectively. The pulse shape is roughly triangular. Note that the point-source approximation was used in this model. The materials initially located above the burial depth can be feasibly ejected by the stress pulse [Melosh, 1985a]. The rarefaction wave can be approximated as the reflected tensile pulse from the mirrored source, which is located at the point $2d$ above the burial depth, due to the zero-pressure boundary condition at the free surface. The compression–decompression history at a given point in the target body can be approximately expressed as a linear superposition of two pulses. If the rarefaction wave can catch up with the propagating compression wave before the compression wave reaches its maximum, then the compressive stress at this point can become much lower than the maximum amplitude of the stress pulse. Hereafter, such wave interference is referred to as "near-surface wave interaction" in this study. The ejection velocity from such a low-compressed region is estimated to be nearly twice the particle velocity in the stress pulse itself, because it is controlled by the pressure gradient between the stress and tensile pulses, and is not the maximum amplitude of the stress pulse [Melosh, 1984, 1985a]. This model provided simple analytical solutions for ejected mass and velocity, and a qualitative understanding of the ejection behavior produced by spallation. However, as noted by Melosh (1984), this approach is not able to calculate the ejection behavior of surface material within a three-fold projectile radius from the impact point, because the assumption of the triangular stress pulse is only valid far from the impact point and the rise time for shock waves are much shorter than that for the triangular stress pulse [e.g., Melosh, 2003]. In addition, the analytical model cannot account for the nonlinear response of realistic geological materials, which results from the nonlinear dependence of density on pressure and thermal pressure in the strongly shocked material. This means that the linear superposition of two pulses is not valid near the impact point.

When an impact-generated stress pulse propagates at a hypersonic speed (i.e., in the presence of a shock wave), the spallation model cannot be applied directly. Nevertheless, the near-surface wave interaction occurs in the case of a shock wave, because the propagation speed of the rarefaction wave in the shocked matter is higher than that of the shock wave. The volume in which downward rarefaction waves from the free surface are able to catch up with the outward-propagating shock wave is known as the "irregular shock reflection region" [e.g., Rosenbaum and Snay, 1956; Kamegai, 1986]. The peak pressure with respect to the distance from the impact point suddenly decreases when the survey line cuts across the irregular shock reflection boundary. We have confirmed that the initial location of the ejecta that meets the MM conditions is in the region as described in Supplementary Materials S1. Although an analytical approach has been proposed to estimate the location of the boundary of the irregular shock reflection boundary [e.g., Rosenbaum and Snay, 1956; Kamegai, 1986], we were unable to analytically obtain the peak pressures of the shocked materials in the irregular shock reflection region. For this reason, hydrodynamic simulations have been conducted to precisely model the interactions between the shock and rarefaction waves, and the intense deformation near the impact point where the point source approximation is not applicable [Head et al., 2002; Artemieva and Ivanov, 2004; Ong and Melosh, 2012].

The presence of thermal pressure is another important difference from the situation considered in the conventional spallation model, when considering the ejection behavior near the impact point. Martian meteorites experienced pressures that ranged from 30 to 50 GPa. The contribution of the thermal pressure component is dominant at this range or higher peak pressures (Supplementary Materials S2). At these peak pressures, the particle velocity behind the shock wave itself cannot be neglected, because it corresponds to a few km/s at a 30 GPa shock compression for granitic and basaltic rocks [e.g., Melosh, 1989]. Following this, fast adiabatic expansion of compressed materials to the free surface is expected to occur to relax the high-pressure state, because adiabatic expansion effectively reduces the thermal pressure (Supplementary Information S2). The adiabatic expansion is expected to drive a further acceleration in the irregular shock reflection region. The resultant particle velocity after the shock-release sequence would mainly cause high-speed material ejection.

We now thermodynamically and hydrodynamically consider the acceleration mechanisms due to a sequence from shock compression to adiabatic expansion. When a shock wave passes into target materials, the materials behind the shock front accelerate in the travelling direction of the wave front. The particle velocity at the peak shock state $u_{\mathrm{pH}}$ can be calculated by the Rankine–Hugoniot relation:

$$P_{\mathrm{peak}} = \rho_0 (C_0 + s u_{\mathrm{pH}}) u_{\mathrm{pH,}} \tag{1}$$

where $P_{\mathrm{peak}}$, $\rho_0$, $C_0$, and $s$ are the peak pressure immediately after the shock passage, the reference density, the bulk sound speed, and a constant, respectively. The strong pressure gradient from the target interior to the free surface relieves the high pressure, resulting in a material flow to the free surface. This is physically the same as subsequent propagation of an expansion wave from the target surface to interior after the rarefaction wave catches up with the shock wave. The shocked materials are accelerated or decelerated during the pressure release. The change in the absolute particle velocity during the pressure release $u_{\mathrm{p\_release}}$ can be calculated using the Riemann invariant along the isentrope, as follows [Melosh, 1989; Kurosawa et al., 2015]:

$$u_{\mathrm{p\_release}} = \int_{\rho_{\mathrm{H}}}^{\rho^*} \frac{C_{\mathrm{R}}}{\rho} d\rho \tag{2}$$

$$C_{\mathrm{R}} = \sqrt{\left.\frac{\partial P}{\partial \rho}\right|_S} \tag{3}$$

where $\rho^*$, $\rho_{\mathrm{H}}$, $C_{\mathrm{R}}$, $\rho$, $P$, and $S$ are the density after the pressure release, density at the peak shock state, speed of sound, density, pressure, and entropy, respectively. All variables apart from $\rho^*$ and $\rho_{\mathrm{H}}$ are values in the expanding material from $\rho_{\mathrm{H}}$ to $\rho^*$. The magnitude of $u_{\mathrm{p\_release}}$ is close to the absolute value of $u_{\mathrm{pH}}$ itself when vaporization after pressure release does not occur, which corresponds to the situation where the integral in Eq. (2) can be approximated by $\rho^* = \rho_0$. The direction of $u_{p\_release}$ is opposite to that of the expansion wave. Since expansion waves reach shocked materials in the irregular shock reflection region after the shock compression, the velocity vector of materials in an excavation flow can be broadly approximated as the sum of the vectors of $u_{pH}$ and $u_{p\_release}$. Thus, the absolute particle velocity after the shock-release sequence strongly depends on the angle between the travel directions of the shock and expansion waves. If the travel direction of an expansion wave is opposite to that of a shock wave, then the absolute value of the particle velocity after the release is ~$2u_{pH}$. This is widely known as the velocity doubling rule at free surfaces [e.g., Melosh, 1989]. Such situations correspond to the rear surface in a planar shock propagation. The ideal maximum for an obtained particle velocity of materials in a condensed phase during shock release is $2u_{pH}$. Consequently, accurate particle velocities depend on local peak pressures and geometric configurations, which determine the angle between shock and expansion waves, and are thus necessary parameters to investigate the ejection behavior near the impact point. It should be noted that the above consideration is only valid for the region where thermal pressure dominantly contributes to peak pressure (i.e., $P_{peak} > 30$ GPa).

Figure 2 shows $2u_{pH}$ as a function of peak pressure for granite, basalt, and dunite. Although granitic and basaltic rocks only just meet the MM conditions, the above considerations of shock release during an impact event indicate that the feasibility of Martian meteorite launch due to impacts is low, as previously suggested by DeCarli et al. (2007). This reflects the fact that $2u_{pH}$ is the maximum ejection velocity from shock release, and that this condition is only met when the traveling direction of the expansion wave is in completely the opposite direction to that of the shock wave. In Section 4.4, we show that the angle between the propagating shock and expansion waves is ~90° in the near-surface spallation region, suggesting that the launch of Martian meteorites by impacts is more difficult than previously thought. As such, it is obvious that additional mechanism(s) for the acceleration of near-surface materials is necessary to launch Martian meteorites. Another objective of this study is to explore these additional mechanism(s), in addition to the effects of artificial viscosity on the near-surface.

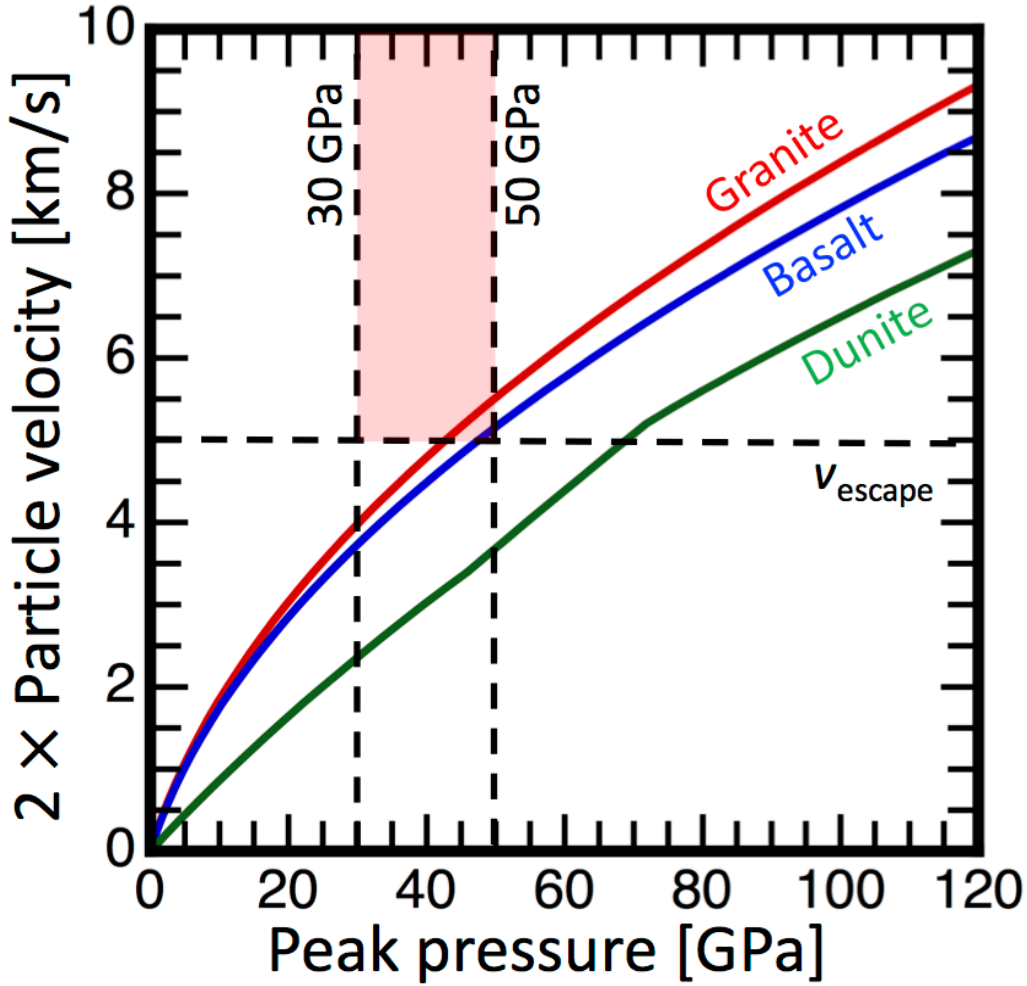

**Figure 2.** Ideal maximum particle velocity $2u_{pH}$ due to shock release as a function of peak pressure. The shock Hugoniot parameters, which are reference density $\rho_0$, bulk sound speed $C_0$, and a constant $s$ for granite, basalt, and dunite, were taken from Melosh (1989): $\rho_0 = 2630$ kg/m$^3$, $C_0 = 3.68$ km/s, and $s = 1.24$ for granite; $\rho_0 = 2860$ kg/m$^3$, $C_0 = 2.60$ km/s, and $s = 1.62$ for basalt. The linear shock-particle velocity relation for dunite is expressed as a piecewise linear function: $\rho_0 = 3320$ kg/m$^3$, $C_0 = 6.6$ km/s, and $s = 0.9$ (<44 GPa), $C_0 = 7.8$ km/s and $s = 0.2$ (44–73 GPa), and $C_0 = 4.4$ km/s and $s = 1.5$ (>73 GPa). The peak pressures at 30 GPa and 50 GPa and the escape velocity of Mars are shown as dotted lines. The red shaded region denotes MM conditions (see main text).

## 3. Hydrocode modeling

We used the 2-D iSALE code and a 3-D SPH code for modeling in this study. A description of the model setups common to both the iSALE and SPH computations is presented in section 3.1. The specific calculation conditions for the iSALE and SPH models are described in sections 3.2. and 3.3, respectively. In section 3.4, the post-analytical procedures used to investigate the nature of the hydrodynamic response near the impact point is explained. It should be noted that we define ejected materials at higher than $u_{pH}$ to be "spalled" in hydrodynamic terms, although the word "spalled" is frequently used to indicate fragments produced due to tensile stress near a free surface.

### 3.1. Model setup

We numerically calculated vertical impact of a spherical projectile with a radius of $R_p = 10$ km onto a flat target. The Tillotson EOS [Tillotson, 1962] was used for both the projectile and target. Table 1 lists the parameters for the Tillotson EOS that we used [Allen, 1967]. Granite was chosen as one of the typical geological materials. Given that the shock impedance of granite is similar to that of basalt (Fig. 2), the result obtained with the EOS for granite is a good approximation for the ejection behavior of basaltic rocks (see also Supplementary Materials S3). It is widely known that the Tillotson EOS oversimplifies the phase diagram of a medium at the expanded region where $\rho < \rho_0$, where $\rho$ and $\rho_0$ are the density and density at the reference state, respectively [e.g., Melosh, 1989]. Nevertheless, this EOS was used in this study because of two reasons: (1) we focused only on ejecta in a condensed phase; (2) the EOS can appropriately treat the thermal pressure and nonlinear dependence of density on cold pressure in the compressed state ($\rho > \rho_0$) [e.g., Tillotson, 1962]. The EOS also nicely reproduce the Hugoniot curve of the shocked medium [Tillotson, 1962]. In both numerical methods, we ignored material strength in order to understand the basic nature of spallation as the result of wave interaction beneath the target surface in hydrodynamic terms. Although the material strength may affect the propagation behavior of shock, rarefaction, and expansion waves [e.g., Bierhaus et al. 2013], the shocked material is likely to be approximated as a perfect fluid, because the focused range of peak pressures in this study is an order of magnitude higher than the Hugoniot elastic limits for typical rocky materials [e.g., Melosh, 1989], which is an upper limit of the compressive strength of damaged rocks [e.g., Collins et al., 2004]. The effects of material strength on ejection behavior is beyond the scope of our study. Gravity was also not considered in the calculations, because gravitational acceleration is expected to be negligible for early stages of the ejection process.

**Table 1.** Input parameters of the Tillotson equations of state [Allen, 1967]

| | |
|---|---|
| Reference density (Mg/m$^3$) | 2.68 |
| Tillotson constant $a$ | 0.5 |
| Tillotson constant $b$ | 1.3 |
| Bulk modulus $A$ (GPa) | 18 |
| Tillotson constant $B$ (GPa) | 18 |
| Tillotson constant $E_o$ (MJ kg$^{-1}$) | 16 |
| Tillotson constant $\alpha$ | 5 |
| Tillotson constant $\beta$ | 5 |
| Specific internal energy for incipient vaporization $E_{iv}$ (MJ kg$^{-1}$) | 3.5 |
| Specific internal energy for complete vaporization $E_{cv}$ (MJ kg$^{-1}$) | 18 |

Although $R_p$ was set to 10 km, we are able to convert our results to any size of impactor, because all hydrodynamic equations can be rewritten in a dimensionless form in cases without gravity and strength [e.g., Johnson and Melosh, 2013]. We set the time to be $t = 0$ at the initial contact between the projectile and target, and calculated the simulations until $t = 1.4\ t_s$. For example, if $D_p$ and $v_{imp}$ are 20 km and 12 km/s, respectively, then $t_s$ becomes 1.7 s. The calculation time is sufficient to investigate the ejection behavior of target materials near the impact point at an ejection velocity of >0.2 $v_{imp}$. The von Neumann–Richtmyer artificial viscosity [von Neumann and Richtmyer, 1950; Monaghan, 1992] was introduced into both the iSALE and SPH calculations, with the same parameters used to capture shock waves and dampen unphysical numerical oscillations behind the shock waves.

*3.2. iSALE 2-D model*

We used the 2-D model of iSALE that is referred to as iSALE-Dellen [Collins et al., 2016] in this study. A cylindrical coordinate system was employed, with directions of $r$ for radius and $z$ for height. Our canonical model divided a spherical projectile into 1000 cells per projectile radius (CPPR). Hereafter, the number of CPPR is referred to as $n_{CPPR}$. The calculated domain was set to 2000 × 3000 cells as a high-resolution zone (HRZ). We also placed extension zones (EZ) on the outside of the HRZ to avoid wave reflections from the calculation boundaries into the HRZ, which would result in a reduction of the total number of computational cells. The size of the cells in the EZ increases as a geometric progression with an extension factor, which is up to 20-times the cell size in the HRZ. We also conducted numerical simulations with different $n_{CPPR}$ values of 125, 250, and 500 to assess the resolution effects, as mentioned in section 1. The impact velocity was fixed at 12 km/s, which is a typical impact velocity onto Mars. The sizes of HRZ and EZ were adjusted appropriately depending on $n_{CPPR}$ to avoid wave reflection and to minimize the computation time. Lagrangian tracer particles were inserted into each computational cell in the HRZ to track material flow through the cells. We stored the temporal variation of the spatial position, pressure, and internal energy of each tracer particle. For computational cost and stability reasons, low density (1 kg/m$^3$) and high-speed cutoffs (10-fold impact velocity) were introduced into the grid-based computations. Given that the density cutoff corresponds to a typical density of air, this does not affect the hydrodynamic motion of the ejecta in the condensed phase. The high-speed cutoff is high enough to investigate the fastest ejecta [Johnson et al., 2014]. In fact, we confirmed that a high-speed cutoff of 50% of our chosen value (five-fold impact velocity) provided similar results to our first models. To constrain the velocity effects on the ejected mass and ascertain if MM conditions were met, we conducted another series of calculations with $n_{CPPR} = 1000$ and impact velocities of 6–21 km/s at a step of 3 km/s. The calculation conditions in the iSALE model setup are summarized in Table 2. It should be noted that our approach using a grid-based

**Table 2.** General setup parameters for the 2-D iSALE calculations

| | |
|---|---|
| Computational geometry | Cylindrical coordinates |
| Number of computational cells in the R direction | 2000 |
| Number of computational cells in the Z direction | 3000 |
| Number of cells for the extension in the R direction | 200 |
| Number of cells for the extension in the Z direction (bottom[a]) | 300 |
| Extension factor | 1.02 |
| Cells per projectile radius (CPPR)[b] | 1000 |
| Grid spacing (m/grid) | 10 |
| Artificial viscosity $a_1$ | 0.24 |
| Artificial viscosity $a_2$ | 1.2 |
| Impact velocity (km/s) | 6, 9, 12, 15, 18, 21 |
| High-speed cutoff | 10-fold impact velocity |
| Low-density cutoff (kg/m$^3$) | 1 |

a. The extension zone in the Z direction was only placed at the bottom of the high-resolution zone.

b. To test for resolution effects, we also performed calculations with CPPR = 125, 250, 500, and 2000 at 12 km/s. The number of computational cells and grid spacing were adjusted depending on CPPR (see Section 3.2).

hydrocode is similar to a previous study by Kamegai (1986), who studied the surface effects on shallow-underwater nuclear explosions using the arbitrary Lagrangian–Eulerian (ALE) code. Although Kamegai (1986) examined the effects of wave interaction beneath the water surface on the pressure field, the author did not focus on changes in particle velocity or the hydrodynamic behavior of geological materials. In our study, we basically followed the approach of Kamegai (1986), but at a higher spatial resolution, using EOS for geological materials and investigating the temporal variations of the particle velocity of materials initially placed near the impact point.

*3.3. Smoothed particle hydrodynamics (3-D-SPH)*

The SPH method [e.g., Lucy, 1977; Monaghan, 1992] is a flexible Lagrangian method of solving hydrodynamic equations that has been widely used for impact simulations in planetary science. The SPH method can easily process large deformations and shock waves. Our numerical code is a 3-D code and is the same as that used by Fukuzaki et al. (2010) and Genda et al. (2015, 2017).

In the SPH calculations, a half-sphere with a radius of two or three times the projectile radius (i.e., 20 or 30 km) was considered as the target. The SPH particles are placed in a 3-D lattice (face-centered cubic) within a sphere of a projectile and a half-sphere of a target. Depending on the numerical resolution, we varied the number of SPH particles. For comparison with the iSALE simulations, the number of SPH particles for the impactor $n_{imp}$ was set to $n_{imp} = 4/3 \; \pi \, n_{CPPR}^{3}$. For example, $n_{imp}$ is 33,510,322 in the case of $n_{CPPR} = 200$. The same resolution for the projectile and target was applied. Thus, the number of SPH particle for the target $n_{tar}$ is 134,241,027 for the case of $n_{CPPR} = 200$. We carried out impact simulations with $n_{CPPR} = 50$, 100, and 200. The calculation conditions in the 3-D-SPH model are summarized in Table 3. Although the SPH calculations were carried out in 3-D coordinates, we only considered vertical impacts in this study to facilitate comparison with the 2-D iSALE model.

*3.4. Data analysis procedures*

Data analyses were conducted based on a particle tracking technique, even when we analyzed the data from the grid-based hydrocode. We extracted the Lagrangian tracer or SPH particles in a condensed phase determined by the Tillotson EOS and used these in the analyses. The condensed phase in the Tillotson EOS is defined by $\rho > \rho_0$ or $\rho < \rho_0$ and $E < E_{iv}$, where $E$ and $E_{iv}$ are the internal energy and the internal energy at incipient vaporization, respectively. In the SPH analysis, we also introduced the same low-density cutoff as used in the iSALE analysis. We analyzed the stored data for each particle, which are the particle velocity vector and pressure. Technically, Lagrangian tracer particles in grid-based hydrocodes do not precisely track fast ejecta. We further consider the accuracy of the tracer tracking in our model in Supplementary Materials

S4.

**Table 3.** General setup parameters for the 3-D SPH calculations

| | |
|---|---|
| Computational geometry | Cartesian coordinates |
| Number of SPH particles in the projectile | 33,510,322 |
| Number of SPH particles in the target | 134,241,027 |
| Corresponding cells per projectile radius (CPPR)[a] | 200 |
| Typical number of neighboring particles | 128 |
| Artificial viscosity $a_1$ | 0.24 |
| Artificial viscosity $a_2$ | 1.2 |
| Impact velocity (km/s) | 12 |
| High-speed cutoff | None |
| Low-density cutoff (kg/m$^3$)[b] | None |

a. We adjusted the number of SPH particles for the projectile to be $4/3 \pi n_{\text{CPPR}}^3$ in order to enable comparison with the iSALE calculations (see text). We also conducted the same calculations with $n_{\text{CPPR}} = 50$ and 100 to examine resolution effects on the impact outcomes in the 3-D-SPH calculations.

b. No density cutoffs were used in the actual SPH calculations; however, we used a low-density cutoff of the same value as used in the iSALE calculations during the post-analysis stage (see Section 3.4.).

## 4. Results

### 4.1. Ejection behavior

Figure 3 shows snapshots of impact simulations from the iSALE and SPH models with different $n_{\text{CPPR}}$ at $t = t_s$. Although motion is calculated in three-dimensions using the SPH code, the data are plotted in cylindrical coordinates. As mentioned in section 3.4, only the particles in the condensed phase ($\rho > \rho_0$ or $\rho < \rho_0$ and $E < E_{\text{iv}}$) are shown in this figure. The particles behind the shock wave are highlighted in different colors depending on the absolute particle velocity. We investigated the effects of artificial viscosity on the ejection behavior using the results over a wide range of $n_{CPPR}$, as discussed in detail in the following section. Our results confirm that the ejection velocities of the top five layers from the target surface for the iSALE 2-D model and the top three layers for the 3-D SPH model do not converge with respect to $n_{CPPR}$, suggesting that they are subjected to artificial viscosity effects. Thus, the particles in these layers are not highlighted in color in Fig. 3 and were not used in subsequent analyses. A hemispherical shock propagation is clearly evident in the computational results. The location of the shock front does not depend on the numerical model or $n_{CPPR}$. In contrast, the location of the leading edge of the ejecta is strongly affected by $n_{CPPR}$, suggesting that the maximum ejection velocity in the numerical calculations is strongly dependent on $n_{CPPR}$. Although the $n_{CPPR}$ values in the 3-D SPH calculations are much lower than in the 2-D iSALE calculations, the spatial distribution of the ejected particles and their particle velocities are similar in both models.

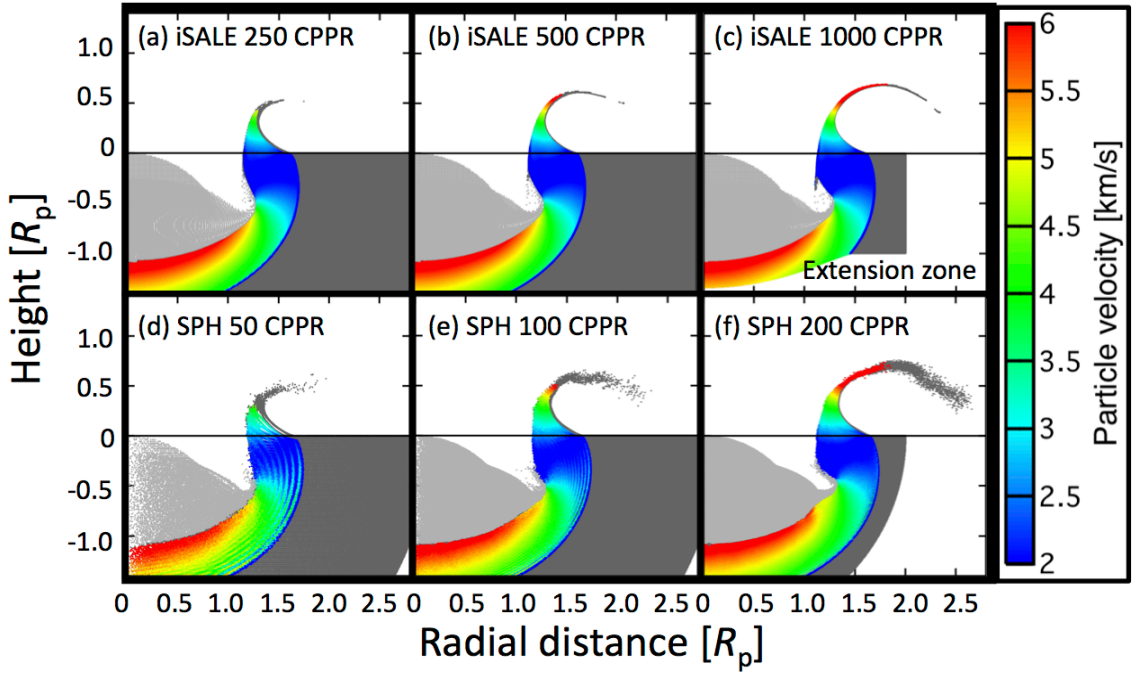

**Figure 3.** Snapshots of the hydrocode calculation results from the iSALE (a–c) and SPH (d–f) codes with different $n_{CPPR}$. We plotted the spatial position of a Lagrangian tracer or SPH particles in a condensed phase at $t = t_s$ in cylindrical coordinates. The $n_{CPPR}$ used in each computation is indicated on the figure. The colors indicate the particle velocities. The particles are initially located in the top five layers for iSALE and top three layers for SPH, and are not highlighted in color (see Section 4.2.). Note that the size of the high-resolution zone (HRZ) in the iSALE calculations with $n_{CPPR}$ = 1000 (c) is smaller than those with $n_{CPPR}$ = 250 (a) and 500 (b). Outside the HRZ and beneath the target surface is the extension zone, which is not a void (see Section 3.2.). Although the outside of the target hemisphere in the SPH is a void, any reflection waves from the boundary between the target materials and void do not interact with the ejected materials in the simulation.

Figure 4 is the same as Fig. 3, except that the former shows the effects of impact velocity on the ejection behavior. The materials moving fast (>5 km/s) have already been ejected at the given times and at any impact velocity. Thus, the end time used in this study ($t = 1.4\,t_s$) is long enough to investigate whether spallation is able to launch Martian meteorites.

### 4.2. Effects of artificial viscosity on ejection behavior

The velocities of iSALE tracer particles and SPH particles vary with time, and some particles are ejected from the target surface. We define the ejection velocity ($v_{eject}$) as the velocity when the particles reach a certain threshold height ($Z_{eject}$) from the target surface. We set $Z_{eject} = 0.1R_p$, because some particles accelerate above the target surface, but this acceleration ceases until the particles are $0.1R_p$ from the target surface. We discuss this acceleration in more detail in section 4.5.

Previous studies have argued that artificial viscosity has a significant effect on ejection behavior, including the relationship between peak pressure $P_{peak}$ and ejection velocity $v_{eject}$ [DeCarli et al., 2007; DeCarli, 2013], as discussed in sections 1 and 2. The shock smearing due to artificial viscosity leads to an artificially low $P_{peak}$ near the target surface. It should be noted that the particle velocities stored on the particles are averaged values of the spatial coverage within the cell sizes, which depend on $n_{CPPR}$. For example, the particles at the top layer for $n_{CPPR} = 125$ and 1000 extend from the free surface down to 80 and 10 m, respectively. Thus, the effects of the difference in the initial depth on the ejection velocity are inevitably included in this resolution test.

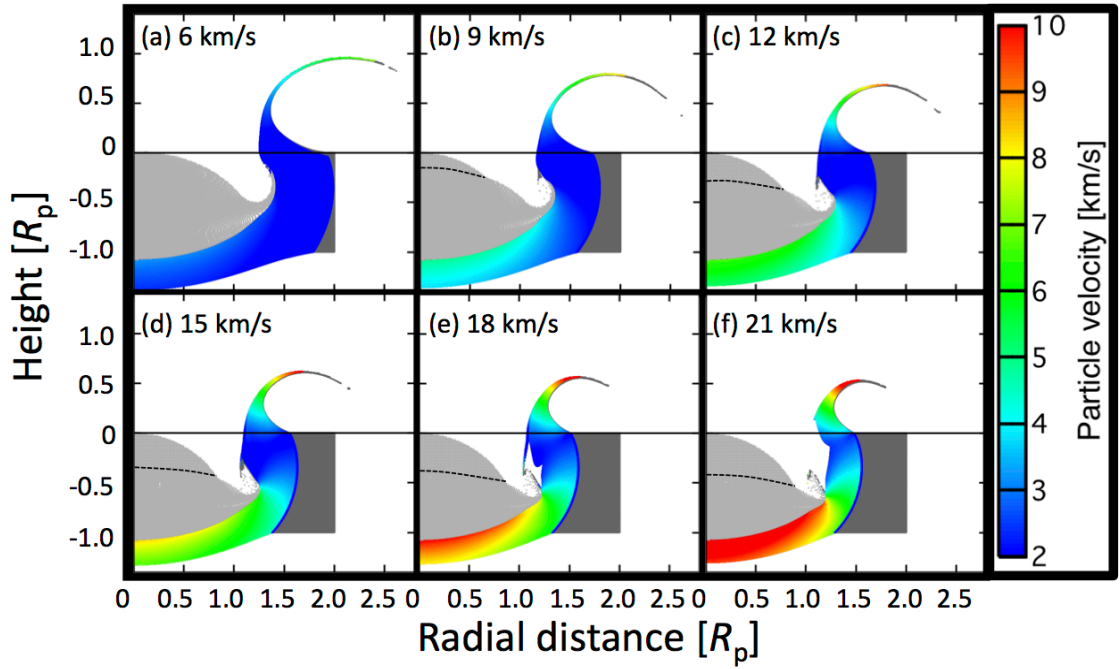

**Figure 4.** Snapshots of the hydrocode calculation results from the iSALE with $n_{CPPR}$ = 1000 showing the velocity effects on ejection behavior. The impact velocities used in the calculations are shown on the figure. The dotted lines in the projectiles indicate the shock fronts in the penetrating projectiles. Note that the real times differ from each other, because $t_s$ depends on the impact velocity.

Nevertheless, the resolution test was designed to remove the particles obtained from unreliable depths in the numerical model and not to explore the detail of shock smearing in the numerical computations. Consequently, we investigated the convergence of the calculation results, such as the $P_{peak}$−$v_{eject}$ relationship, with respect to spatial resolution. To investigate the validity of our nominal models, we conducted an additional iSALE simulation at a higher spatial resolution of $n_{CPPR}$ = 2000. In this case, the end time was set to $t$ = 0.6 $t_s$ to reduce the computational cost. This end-time for the simulation with $n_{CPPR}$ = 2000 was chosen because the fast moving ejecta (>5 km/s) is launched prior to $t$ = 0.6 $t_s$.

Figure 5 shows the ejection velocities of the tracer particles initially located at the top layer of the target as a function of the peak pressures. The ejection velocity at a given peak pressure does not converge with respect to $n_{CPPR}$, suggesting that the obtained ejection velocities are not reliable, as noted previously [DeCarli et al., 2007; DeCarli, 2013]. Figure 6 is the same as Fig. 5, except that it shows the initial depth of the ejected particles. The initial depth is 0.5%–1.0% of $R_p$ from the target surface, which corresponds to the 6–10th layers, 11–20th layers, and 4th layer from the target surface with $n_{CPPR} = 1000$ and 2000 for iSALE, and with $n_{CPPR} = 200$ for SPH, respectively. Our results confirm that the $P_{peak}$–$v_{eject}$ relationship for this depth converges into a similar region on this plot for at least $P_{peak} < 70$ GPa, suggesting that layers deeper than the 5th layer from the target surface are not subjected to numerical artifacts in the iSALE computation. Based on the similar test against to the SPH results, we confirmed that the particles from deeper than 3rd layer from the target surface converge into the same region on the $P_{peak}$-$v_{ej}$ plot (See Supplementary Materials S5.)

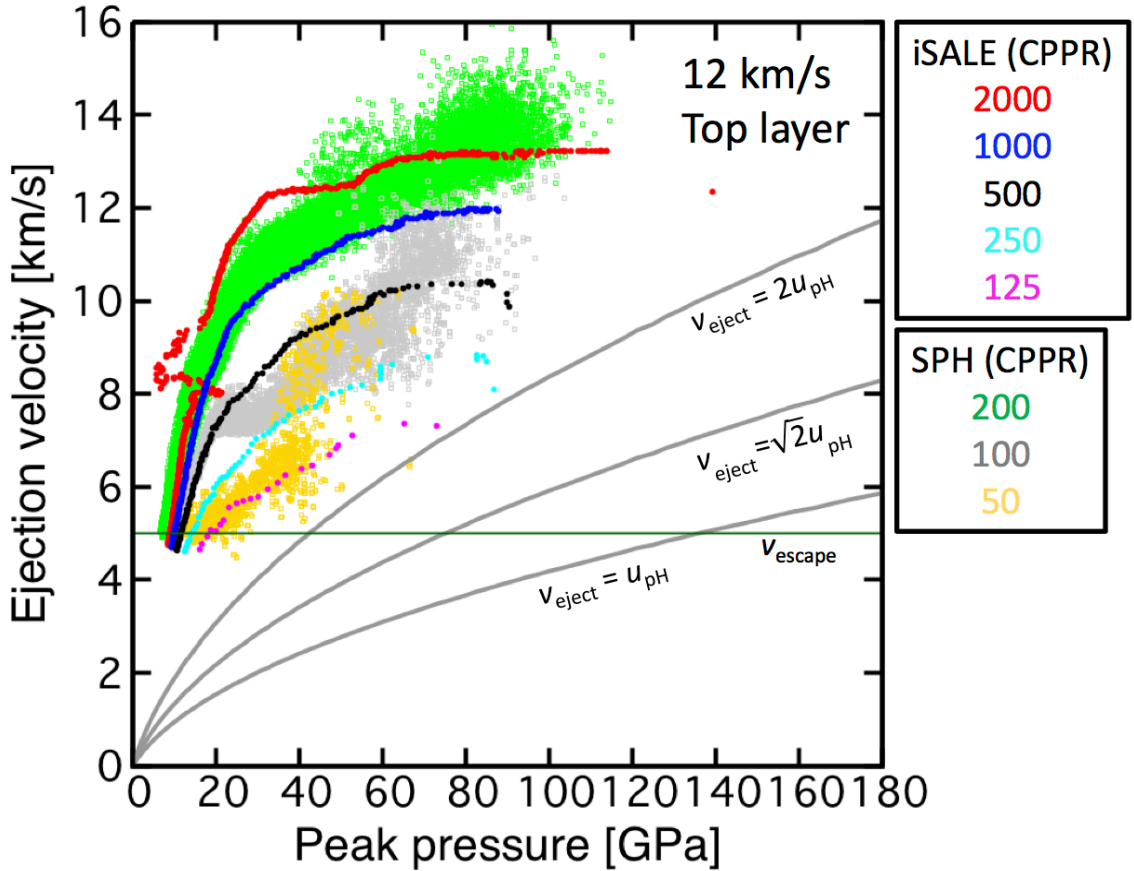

**Figure 5.** Ejection velocity as a function of peak pressure. Only the particles initially located in the top layer of the target are shown here. The escape velocity of Mars $v_{escape}$ and the curves for $v_{eject} = u_{pH}$, $\sqrt{2}u_{pH}$, and $2u_{pH}$ are also shown as guides (See Section 2). The number of plotted particles increases at a larger $n_{CPPR}$, and in the 3-D SPH model is much larger than in the 2-D iSALE model.

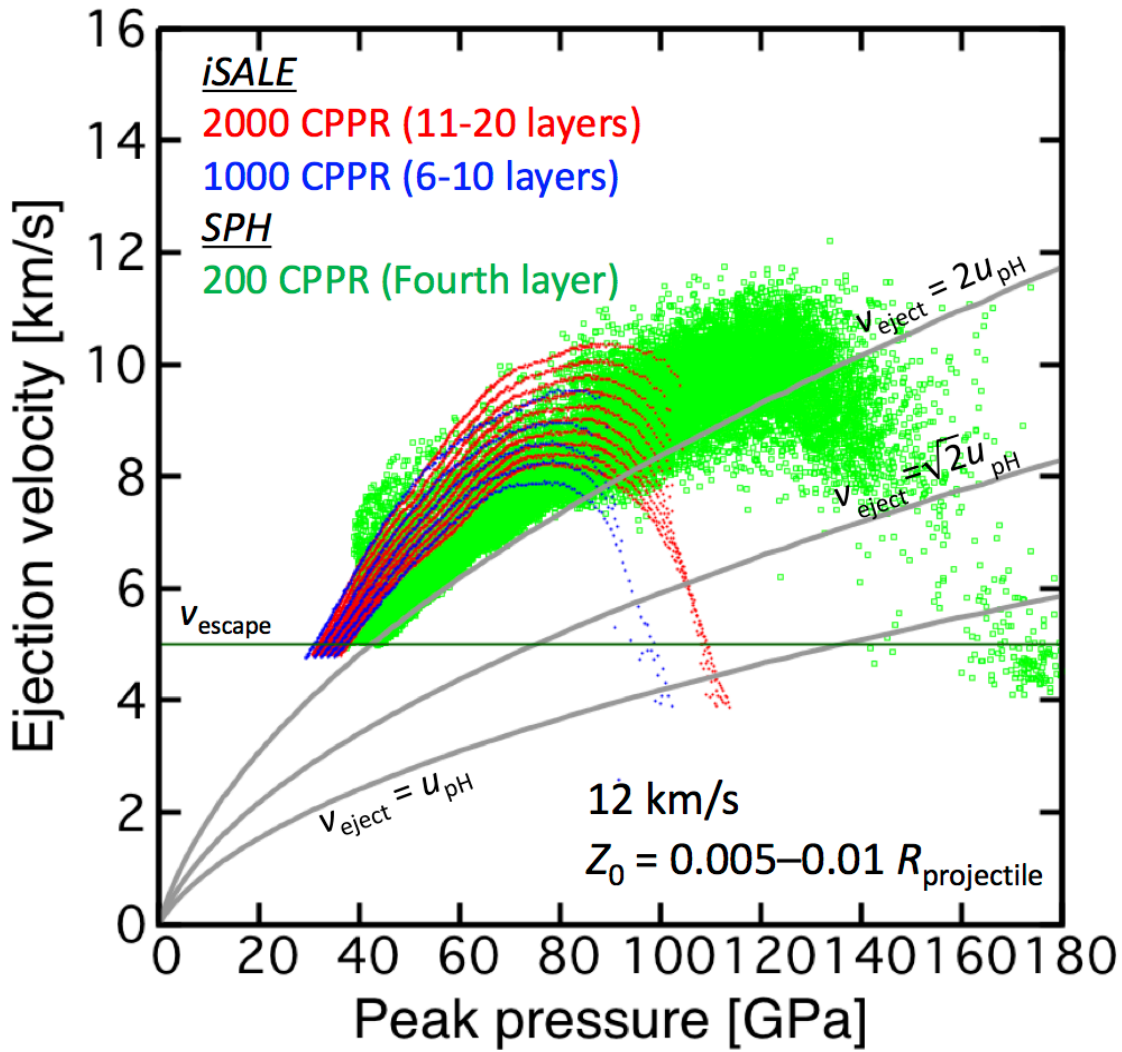

Figure 6. Ejection velocity as a function of peak pressure, but only showing the particles initially located at a depth of 0.5%–1.0% of the projectile radius. The corresponding orders of the layers counted from the target surface are indicated on the figure. Only the 2-D iSALE results with $n_{CPPR}$ = 1000 and 2000 and the 3-D SPH results with $n_{CPPR}$ = 200 are shown here.

To discuss the inter-code variability, we refer to Fig. 6, where $P_{peak}$ at >90 GPa is different for each model. For example, $v_{eject}$ with $n_{CPPR}$ = 2000 is twice that of when $n_{CPPR} = 1000$ at $P_{peak} = 100$ GPa, even for the same iSALE model. In addition, there is no particle with $P_{peak} > 120$ GPa in the iSALE results, which contrasts with the SPH results where $P_{peak}$ reached up to 180 GPa. This difference at $P_{peak} > 100$ GPa may be due to the artificial oscillations of pressure behind the shock front in the SPH simulation, which is clearly observed in Fig. 3d. The coefficients of the artificial viscosity employed in this study are smaller than them used in Genda et al. (2017). Nevertheless, the $P_{peak}$–$v_{eject}$ relationship at $P_{peak} < 80$ GPa from the three results, which includes the MM conditions, is in good agreement. Consequently, the top five layers for the iSALE model and the top three layers for the SPH model were not included in the following analysis.

### 4.3. Relationship between peak pressure and ejection velocity

In this section, we present the ejection velocity $v_{eject}$ as a function of the peak pressure $P_{peak}$ of each particle. Figure 7 shows the ejection velocities of particles initially located at different depths as a function of peak pressure during vertical impacts at 6 km/s (Fig. 7a) and 12 km/s (Fig. 7b). Only the iSALE results with $n_{CPPR} = 1000$ are shown. The curves for $v_{eject} = u_{pH}$, $\sqrt{2}u_{pH}$, and $2u_{pH}$ are also shown as guides as mentioned in Section 2. The first curve corresponds to the Rankine–Hugoniot relationship between pressure and particle velocity calculated by Eq. (1). The second curve is the maximum particle velocity obtained from considerations of the geometric interaction between shock and expansion waves, as discussed in the next section. The final curve corresponds to the ideal maximum value of the resultant particle velocity after shock release, as discussed in section 2. The red shaded region indicates the MM conditions. Figure 7 clearly shows that the ejection velocity is able to exceed the $2u_{pH}$ value, in contrast to the doubts raised in previous studies based on shock physics [DeCarli et al., 2007; DeCarli, 2013]. The particles do not appear to have any clear correlation with respect to $P_{peak}$ (Fig. 7), implying that the material ejection from the irregular shock reflection region are affected by not only $P_{peak}$, but also the geometric configuration as discussed in Section 2. To clarify the acceleration mechanism, the ejected particles were divided into three groups: (1) $v_{eject} < u_{pH}$, (2) $u_{pH} < v_{eject} < \sqrt{2}u_{pH}$, and (3) $v_{eject} > \sqrt{2}u_{pH}$. Group 1 is possibly ejecta due to normal excavation [e.g., Melosh, 1985b]. The production processes for groups 2 and 3 are discussed in sections 4.4 and 4.5, respectively. In this study, we refer to the ejected materials in groups 2 or 3 as spalled materials.

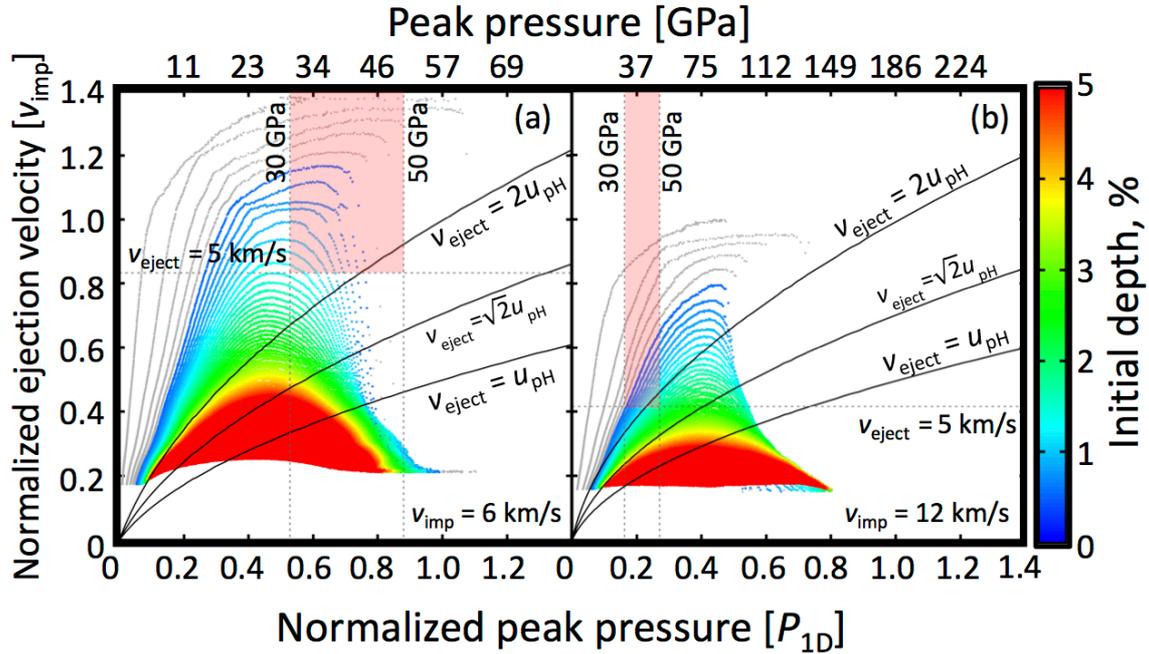

**Figure 7.** Ejection velocities of the particles initially located at different depths as a function of the peak pressures at 6 km/s (a) and 12 km/s (b). The ejection velocities and peak pressures are normalized to the impact velocities and the pressures $P_{1D}$ calculated using 1D impedance match solutions and Eq. (1), respectively. The red shaded region represents the MM conditions as in Fig. 2. The values on the upper X-axis indicate actual peak pressures. The colors indicate the initial depth expressed as a percentage of the projectile radius.

We discussed the limitation of the Melosh spallation model in the case of the presence of the shock wave in Section 2. Nevertheless, we conducted an additional analysis based on the Melosh spallation model in Supplementary Information S6. The relationship between the pressure without the surface effect $P_{free}$ and $v_{ej}$ is presented (See Figure S8). If the spallation model can be applied to the case of the presence of shock wave, the ejection velocity does not exceed $2u_{pH}$, which is calculated at $P_{free}$ by Eq. (1). Our test clearly shows that there are the tracer particles, which are ejected higher than $2u_{\mathrm{pH}}$, although the number of such tracers are relatively small.

### 4.4. Propagation of the shock and expansion waves

We examined the propagation directions of both the shock and expansion waves to understand the effects of the geometric configuration on the $v_{\mathrm{eject}}$–$P_{\mathrm{peak}}$ relationship. Material behind a shock wave is accelerated in the same direction as the propagation direction of the wave. In contrast, a material during pressure release is accelerated in the opposite direction to the propagation direction of the expansion wave. Thus, it is important to determine the angle between the propagating shock and expansion waves as discussed in Section 2.

Figure 8a shows isochrones of the shock front and expansion wave front plotted as the initial position of each particle. The iSALE result with $n_{\mathrm{CPPR}} = 1000$ for an impact at 12 km/s is shown. The arrival time of the shock front is calculated as the time when the pressure of each particle exceeds 1 GPa from the undisturbed state. The arrival time of the expansion wave front is calculated to be the time when the pressures of the shocked particles decrease to one-third of the peak pressures. A shock wave propagates into the target with a hemispherical shape from the impact point. The distance between the shock fronts at two neighboring sites gradually decreases with time, suggesting decaying shock propagation. An expansion wave was generated at around 0.4 $t_{\mathrm{s}}$ and follows the shock front with a near-triangular shape with a central point at around 0.8 $R_{\mathrm{p}}$, where it is slightly inside the edge of the projectile footprint at the target surface.

Figure 8b shows a close-up view of the contours near the target surface of the edge of the projectile footprint. The isochrones of the shock front slightly deviate from the vertical due to a stronger shock decay at the near-surface. The isochrones of the expansion wave front also deviate from the horizontal, because the shocked materials initially located at a point closer to the impact point recover from the high pressure at an earlier time than those further from the impact point. As a result, the angle between the shock and expansion waves is ~90°.

Figure 8c shows schematics of the particle velocity vectors after shock release at different points (A and B) shown in Fig. 8a. At point A, located in a shallow position, the angle between the traveling directions of the shock and expansion wave fronts becomes ~90°. Therefore, the resultant maximum particle velocity immediately after the shock-release sequence is expected to be ~$\sqrt{2}u_{\text{pH}}$. In contrast, at point B, located at a deeper position, the angle between the traveling direction of the shock and expansion wave fronts is relatively small. Here the resultant particle velocity is likely to be close to the $u_{\text{p\_res}}$ defined in Section 2, which is typically much less than the impact velocity [e.g., Melosh, 1985b], resulting in normal excavation. Note that the threshold to calculate the location of the expansion wave front cannot be defined as a certain value, because the pressure of shocked materials gradually decreases with time until it becomes zero. An arbitrary choice for the threshold is necessary, which we set to be $(1/3)P_{\text{peak}}$. Figure 9 shows the velocity and acceleration vectors on the expansion wave front at $t = 0.5\ t_s$, along with the all the tracer locations colored depending on temporal pressure. Only the iSALE results with $n_{\text{CPPR}} = 1000$ at 12 km/s are shown. The black line indicates the location of the expansion wave front, where the temporal pressure is $(1/3)P_{\text{peak}}$ at this time. The directions of acceleration vectors, which correspond to the direction of the local pressure gradients, correlate with the expansion wave front. This figure clearly shows that the situation of the point A shown schematically in Fig. 8c actually occurred in the numerical model.

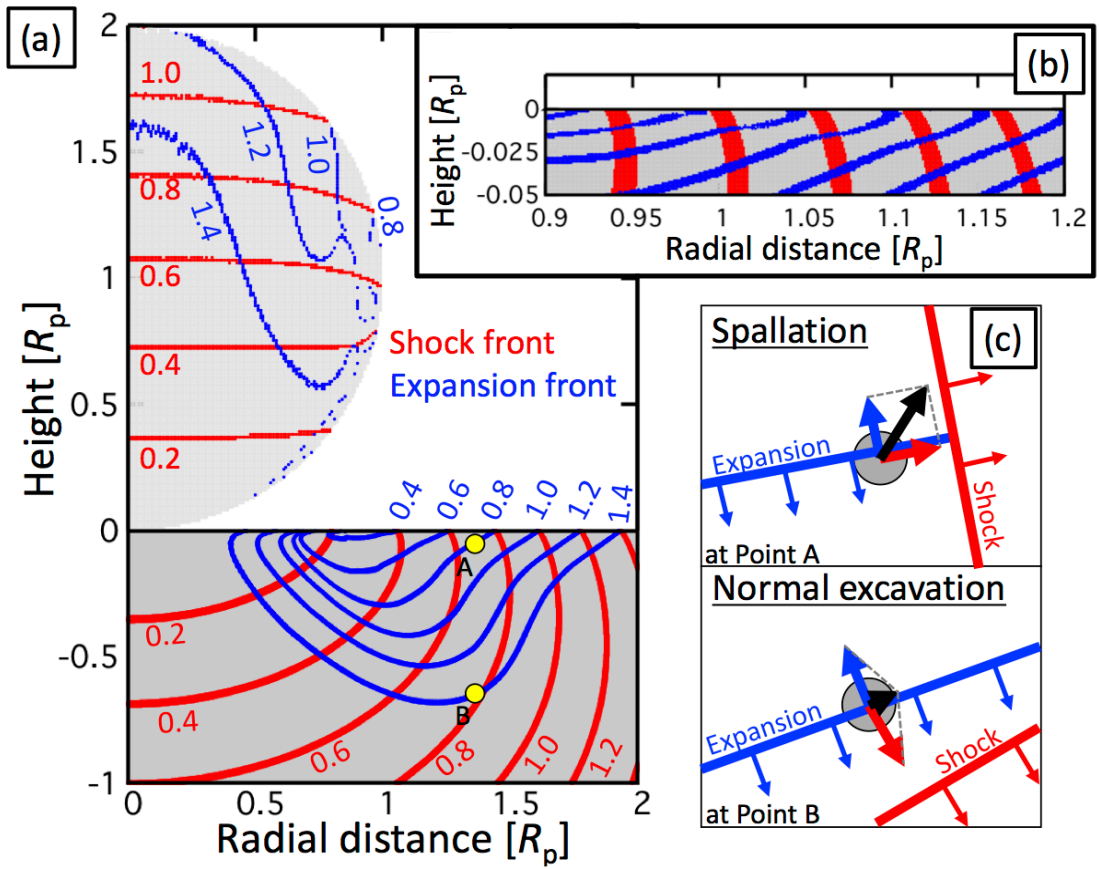

**Figure 8.** (a) Isochrones of the shock front (red) and expansion wave front (blue) that are plotted in the initial position of each particle. The corresponding scaled time of each isochrone is indicated beside the lines. Only the iSALE result with $n_{\text{CPPR}} = 1000$ for an impact at 12 km/s is shown here. (b) Close-up view of the isochrones near the target surface around the edge of the projectile footprint. (c) Schematic diagram of the resultant velocity vectors after shock release at different points (A and B) indicated in (a).

Given the analysis described in this subsection, we propose that the "actual" maximum particle velocity immediately after a shock-release sequence in the case of 2-D shock propagation is $\sqrt{2}u_{\text{pH}}$, rather than $2u_{\text{pH}}$ as conventionally used in previous studies [e.g., DeCarli et al., 2007]. Although the $P_{\text{peak}}$–$v_{\text{ej}}$ relationship does not converge onto the $\sqrt{2}u_{\text{pH}}$ line, the group 2 ejected material defined in Section 4.3 can be explained by near-surface interaction between the shock and expansion waves.

### 4.5. Late-stage acceleration

According to the simple physical considerations in the previous section, we found that the maximum particle velocity of material initially located near the target surface is expected to be $\sim\sqrt{2}u_{\text{pH}}$ after shock release. However, some particles near the target surface are accelerated to $>\sqrt{2}u_{\text{pH}}$ (Fig. 7). Although particles from the uppermost layers of the target would exceed $\sqrt{2}u_{\text{pH}}$ due to the artificially low $P_{\text{peak}}$ resulting from shock smearing, some particles below this are actually accelerated to $>\sqrt{2}u_{\text{pH}}$ due to an unknown mechanism. In this section, we investigate this unknown mechanism that accelerates group 3 ejecta to velocities $>\sqrt{2}u_{\text{pH}}$.

Figure 10 shows snapshots of close-up views around the edge of the projectile footprint. Only the iSALE result with $n_{CPPR} = 1000$ for an impact at 12 km/s is shown. The pressure distribution, trajectories of the six selected tracers with velocity vectors, and tracer overlays at the same depth are also shown. We found that the temporal pressure (not peak pressure) of the tracer particles forming the root of the ejecta curtain is still >10 GPa, in spite of their height above the target surface. This causes further acceleration due to the pressure gradient from the center to the outside of the ejecta curtain.

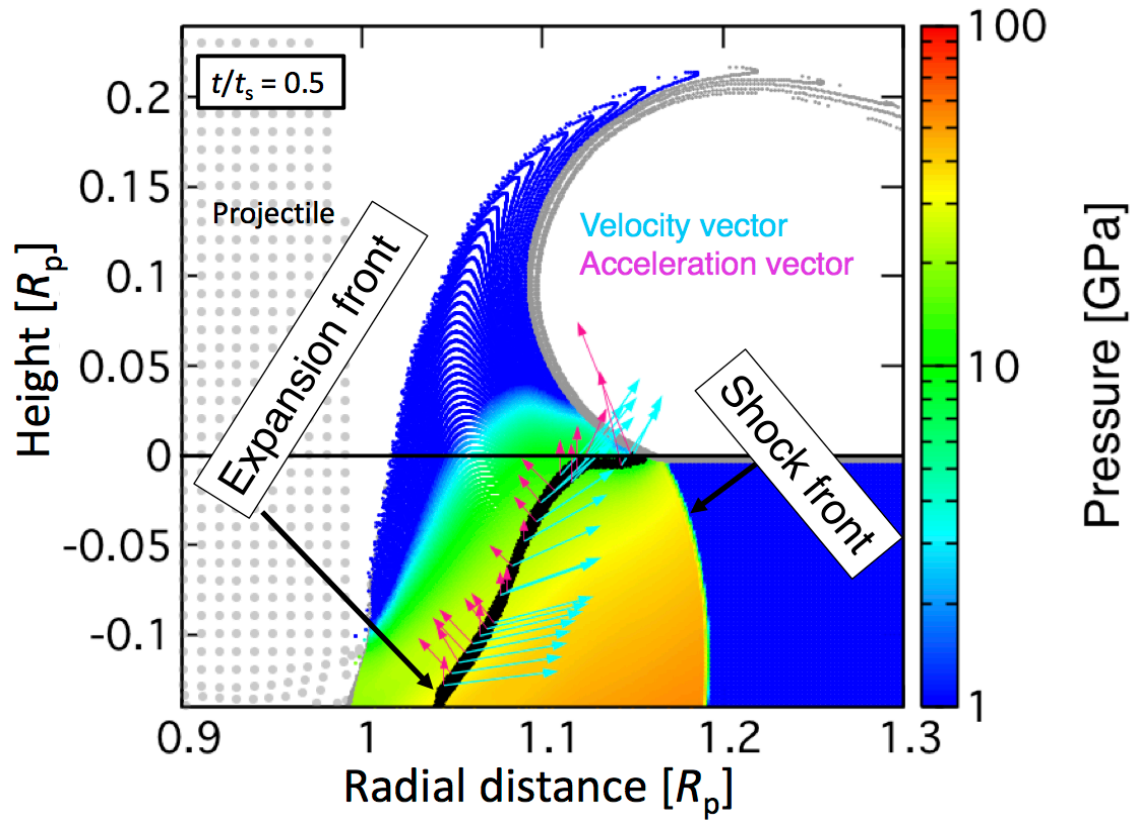

**Figure 9.** A snapshot of close-up views around the edge of the projectile footprint at 0.5 $t_s$, along with velocity (greenish blue arrows) and acceleration vectors (dark pink arrows) on the expansion wave front (black line). Only the iSALE result with $n_{CPPR} = 1000$ for an impact at 12 km/s is shown. The temporal pressures (i.e., not peak pressure) of tracer particles are also shown highlighted in color, except for the tracers initially located in the top five layers (Section 4.2.). The definition of the location of the expansion wave front is described in Section 4.4.

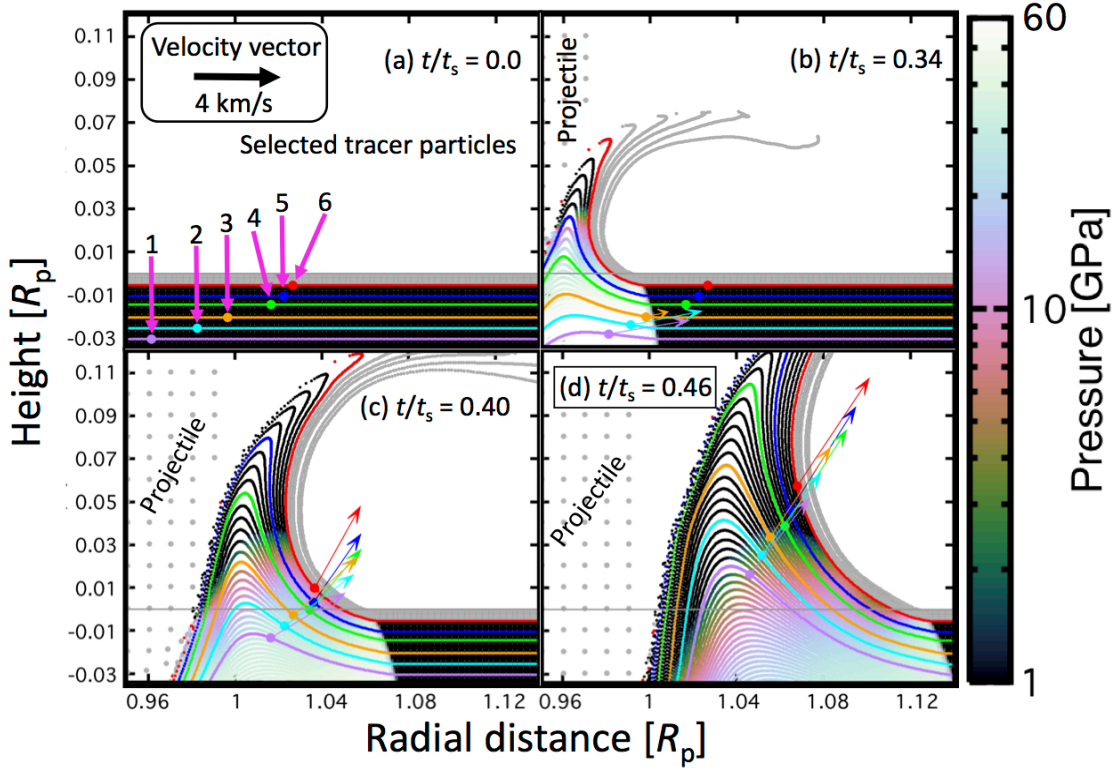

**Figure 10.** Snapshots of close-up views around the edge of the projectile footprint. The scaled time is shown. The temporal pressures (i.e., not peak pressure) of tracer particles are highlighted in color, except for the tracers initially located in the top five layers (Section 4.2.). The six selected tracers with velocity vectors and the tracer overlays at the same depth as the selected tracers are also shown. The initial depths of tracers #1– #6 are described in the main text.

The six tracers were chosen because they mostly tracked the same trajectory (Fig. 10d). The velocity vectors are oriented in the same direction as the propagating shock wave (Fig. 10b). However, the direction of the velocity vectors gradually direct upward (Fig. 10c), indicating that the expansion waves reach the six tracers from above. Figure 11a shows the temporal evolution of the particle velocities (solid lines) and pressure (dotted lines) of selected tracer particles numbered in Fig. 10a. Tracers #1 to #6 were initially located at the 31st, 26th, 21st, 15th, 11th, and 6th layers from the target surface, respectively. We confirmed that the rise times of the pressures are much shorter than $t_s$ at the range of $P_{peak}$ of this study. The peak pressures of the tracers correlate well with the initial distances from the impact point. The particle velocities immediately after the passage of the shock wave are close to $u_{pH}$ for all six tracer particles. The particle velocities then increase with time as the pressures decreases due to the arrival of expansion waves. It is notable that the terminal particle velocities for tracers #4, #5, and #6 exceed $\sqrt{2}u_{pH}$ and meet the MM condition. Although the $u_{pH}$ of tracer #1 is highest, the terminal particle velocity, or $v_{eject}$, of tracer #1 is the lowest, suggesting that $v_{eject}$ strongly depends on the initial position. Figure 11b shows the temporal variation of the distance from the impact point. The colors indicate the temporal pressure. We found that the inter-particle distance between the six tracers, which mostly have the same trajectory, becomes a minimum at around $t = 0.4 \, t_s$, indicating a material pileup. At this time, the tracers are located around the ground level (Fig. 10c), plateaus in temporal pressures at 10–30 GPa are obvious (dotted lines in Fig. 11a), and the magnitude relationship of the particle velocity is reversed (solid lines in Fig. 11a). Following this, the tracers moved in uniform linear motions. Figure 12 shows the temporal variation of the particle velocity and temporal pressure of tracer #6. The radial ($u_{pR}$, grey) and vertical ($u_{pZ}$, black) components of the particle velocity and direction of the velocity vector measured from the horizontal in a counter-clockwise fashion (the color bar) are also shown. A sudden rise in pressure up to ~40 GPa indicates the arrival of the shock wave. The shock incidence angle of this tracer is 20–30°. Since the rarefaction wave is expected to have already reached the initial location of the tracer, the temporal pressure starts to decrease down to ~10 GPa within ~0.02 $t_s$ immediately after the shock wave arrival. During the pressure release, $u_{pR}$ slightly decreases and $u_{pZ}$ increases up to ~3 km/s, resulting in a change in the direction of the particle velocity from 20–30° to 60–70°. The resultant particle velocity approaches $\sqrt{2}u_{pH}$. This behavior is consistent with the predicted change due to the shock-release sequence discussed in Section 2. Following this, the pressure stays at ~10 GPa from 0.39 $t_s$ to 0.41 $t_s$. During this phase of constant pressure, both $u_{pR}$ and $u_{pZ}$ gradually increase. The pressure then further decreases down to zero at ~0.45 $t_s$, leading to the uniform linear motion. The height of the tracer at this time is ~0.05 $R_p$. The terminal particle velocity exceeds $2u_{pH}$.

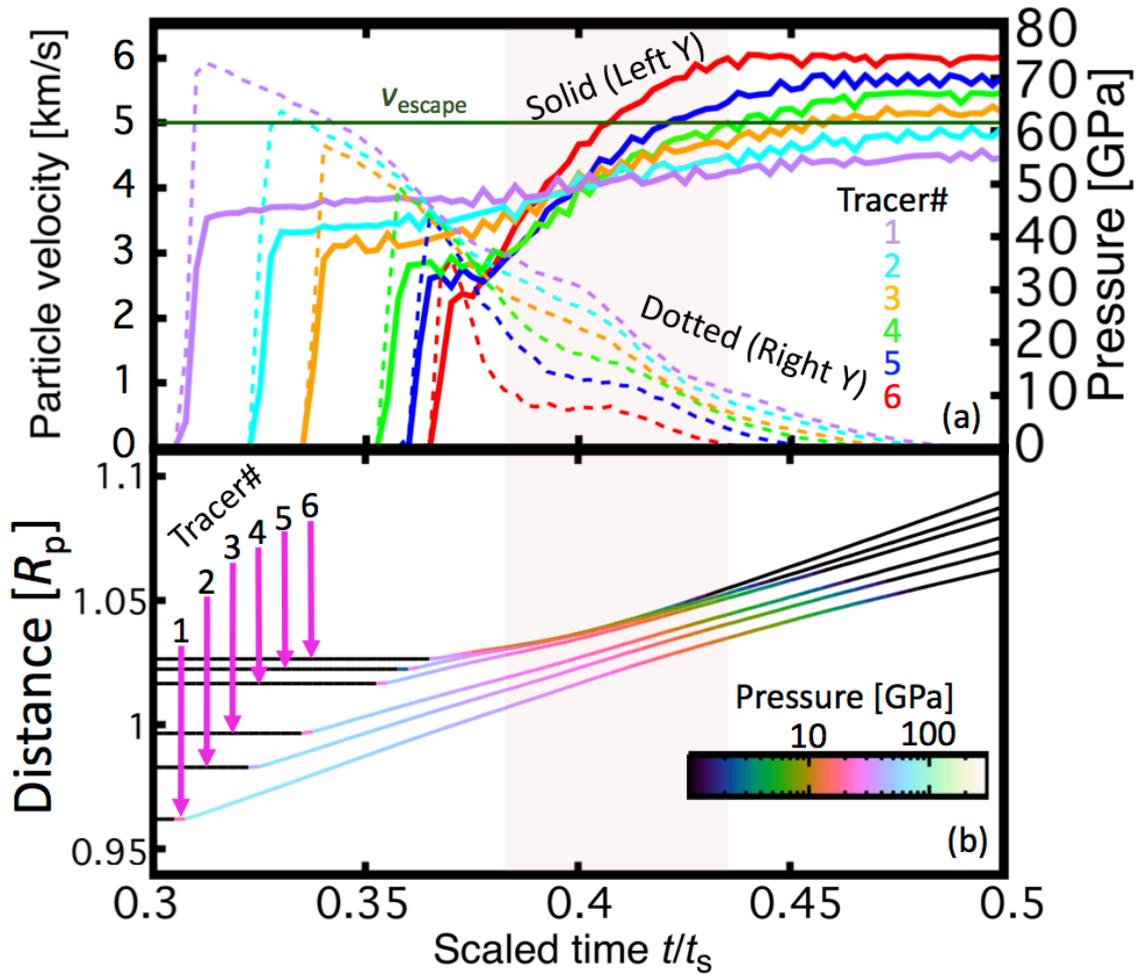

**Figure 11.** (a) Temporal variations in particle velocity and pressure for the selected tracer particles numbered in Fig. 10a. The left and right Y-axes indicate particle velocity (solid lines) and pressure (dotted lines), respectively. The escape velocity of Mars ($v_{escape}$) is shown as a horizontal green line. The color corresponds to each tracer number. (b) Temporal variations in the distance from the impact point for the selected tracers. The colors indicate the temporal pressure. The red shaded region indicates the timing of the late-stage acceleration (Section 4.5).

At $t = 0.4 \, t_s$ (Fig. 10c), tracers #4, #5, and #6 are in the root of the ejecta curtain and are accelerated upward and outward due to the pressure gradient from the center of the root of the ejecta curtain to the free surface. The pressure gradient is expected to be produced by pileup in the ejection flow, which would originate from the difference in $u_{pH}$ that depends on the initial distance from the impact point, which in turn resulted from the decaying shock propagation (Fig. 11a). This indicates that excavation flow at an early stage is a compressible flow, despite the fact that excavation flow at later stages can be approximated as an incompressible flow [e.g., Maxwell, 1977].

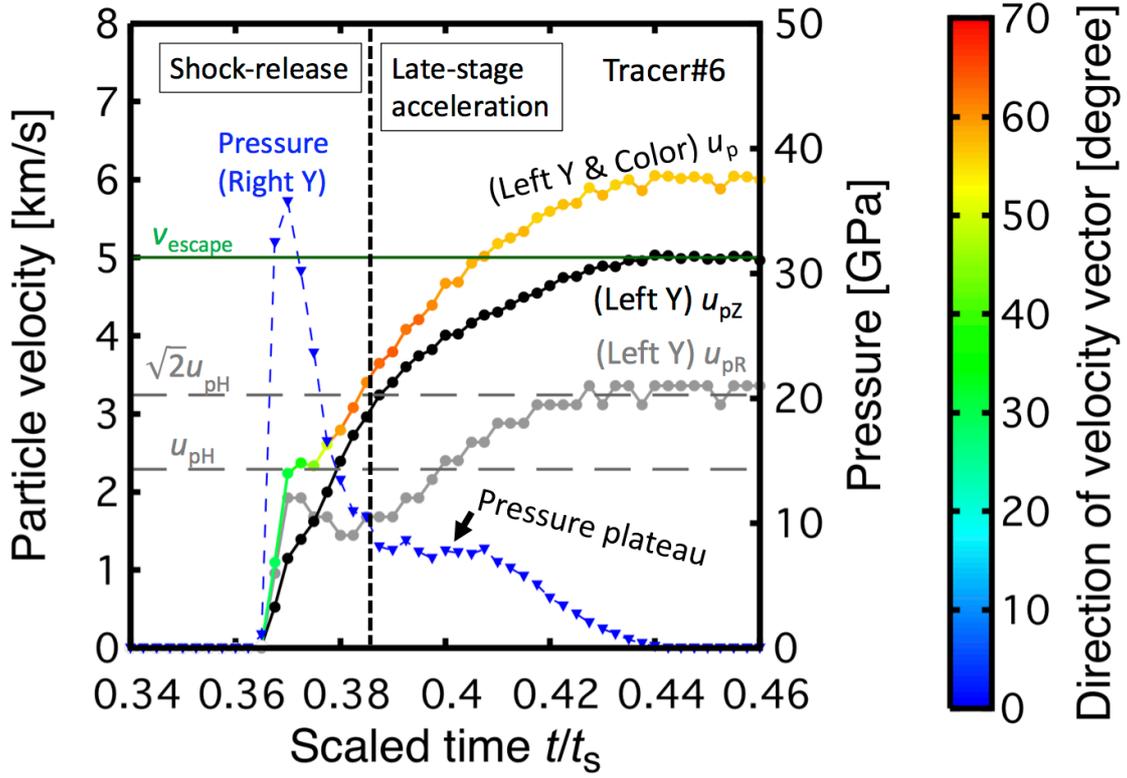

**Figure 12.** Temporal variations in particle velocity and pressure for the tracer #6. The left and right Y-axes indicate particle velocity (solid lines) and pressure (blue dotted lines), respectively. The horizontal and vertical components of the particle velocity are also shown as grey and black lines, respectively. The colors on the particle velocity indicate the direction of the velocity vector. Two horizontal dotted lines labeled $u_{pH}$ and $\sqrt{2}u_{pH}$ correspond to the particle velocities immediately after shock arrival and maximum particle velocities after pressure release, respectively (Section 4.4). The vertical dotted line indicates the time when the adiabatic expansion ceases and material pileup in the ejection flow becomes obvious.

Given that the pressure of the outer edge of the ejecta curtain is approximately zero, the magnitude of the acceleration $a_{late}$ is roughly expressed as follows:

$$a_{\text{late}} = \frac{1}{\rho} \frac{\partial P}{\partial r} = \frac{P_{\text{root}}}{\rho l} \qquad (4)$$

where $\rho \sim 3000$ kg/m$^3$, $P_{\text{root}}$, and $l$ are the density of the materials in the ejection flow, pressure in the root of the ejecta curtain, and the thickness of the curtain, respectively. The velocity increase $\Delta v$ is described as $\Delta v = a_{\text{late}} \Delta t = \frac{P_{\text{root}}}{\rho l} \Delta t$, where $\Delta t$ is the duration of the compressible flow. If we introduce two free parameters $\alpha$ and $\beta$ to be $l = \alpha R_{\text{p}}$ and $\Delta t = \beta t_s = 2\beta R_{\text{p}}/v_{\text{imp}}$ into the above approximation, then $\Delta v$ can be rewritten as

$$\Delta v = \frac{P_{\text{root}}}{\rho} \frac{2\beta}{\alpha v_{\text{imp}}} \qquad (5)$$

Our calculations show that $P_{\text{root}} \sim 10$ GPa, $\alpha \sim 0.05$, and $\beta \sim 0.05$ (Figs. 11 and 12), resulting in $\Delta v \sim 5$ km/s at an impact velocity of 12 km/s. This $\Delta v$ is the same order of magnitude as the observed $\Delta v$, which is calculated from the difference between $v_{\text{eject}}$ and $\sqrt{2} u_{\text{pH}}$. Consequently, the pressure gradient in the ejecta curtain is large enough to accelerate up to $> \sqrt{2} u_{\text{pH}}$ during the ejection. Although the acceleration by the pressure gradient was originally proposed in the spallation model of Melosh (1984), the generation mechanism of the pressure gradient is quite different. The pressure gradient in Melosh (1984) is generated between two triangular pulses, which are a preceded stress and a subsequent tensile ones. In contrast, we found that the material pileup at the root of the ejecta curtain, which comes from the difference in $u_{\text{pH}}$ depending on their initial positions, produce the pressure gradient directed to the upward and outward. Hereafter, we refer the newly described acceleration mechanism as "late-stage acceleration". The acceleration effectively occurs above the ground up to $\sim 0.05$ Rp. Thus, we adopted an ejection height threshold $Z_{\text{eject}} = 0.1 \, R_{\text{p}}$ above the target surface instead of the ground ($Z_{\text{eject}} = 0$), as mentioned in Section 4.2.

### 4.6. Spalled mass

In this section, we consider the high-speed spalled materials ejected up to $t = t_s$. Material ejected at one-third of the impact velocity is launched within $1t_s$ under the iSALE calculation conditions. Figure 13a and b show the resolution effects on the spalled mass (groups 2 and 3), and the mass of materials that experienced late-stage acceleration (group 3 only) during a vertical impact at 12 km/s, respectively. The mass has a positive correlation with $n_{CPPR}$ regardless of the numerical model used. We find that the spalled mass calculated from the 2-D iSALE model with $n_{CPPR} \geq 250$ and from the 3-D SPH model with $n_{CPPR} \geq 100$ is a linear function of the inverse of $n_{CPPR}$ (Fig. 13a). The spalled mass at $n_{CPPR} = \infty$ can be estimated by extrapolation of the straight line to the Y-axis. The extrapolated results for the iSALE and SPH models are in excellent agreement, strongly supporting the validity of our numerical model. The spalled mass during a vertical impact at 12 km/s reaches ~10 wt% of the projectile mass, which is one to two orders of magnitude larger than the jetted mass. Although the masses ejected at $>\sqrt{2}u_{pH}$ with $n_{CPPR} = \infty$ calculated from the two different models are not in full agreement, the difference is only a factor of 1.4.

### 4.7. Launch position and total mass of ejected materials meeting Martian meteorite conditions

In previous sections, we demonstrated that spallation is able to produce fast ejecta moving at velocities $>2u_{pH}$ due to late-stage acceleration near the surface. Here, we examine the initial location of ejected materials that meet MM conditions, where the peak pressure $P_{peak} = 30$–$50$ GPa and the ejection velocity $v_{eject} > 5$ km/s. Hereafter, such ejecta is referred to as "MM ejecta". We also examine the total mass of MM ejecta.

Figure 14a shows the initial position of ejecta launched prior to $1.4t_s$ colored depending on the ejection velocity. The isobaric lines for given peak pressures are also shown. It is clear that high-velocity ejecta are limited to material initially located in the shallow part of the target ($<0.2 \, R_p$ from the target surface). Figure 14b shows the initial position of the MM ejecta. The three boundaries constrained by the range of peak pressure and escape velocity are indicated. The initial depth of the MM ejecta is within 2% of $R_p$.

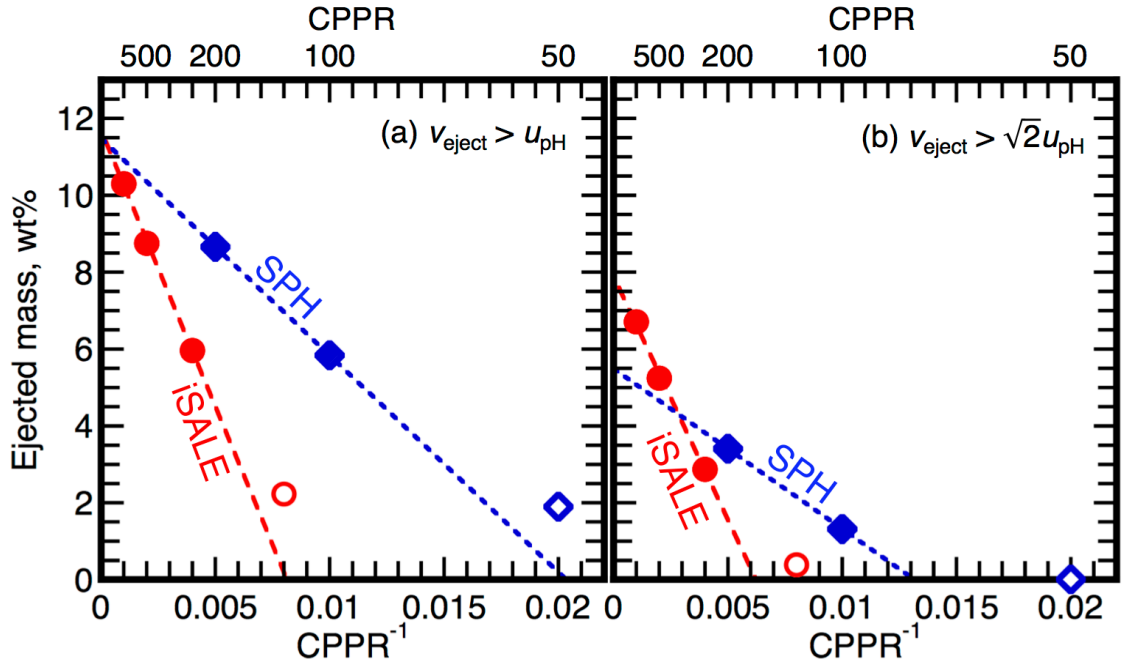

**Figure 13.** Ejected mass as a function of the inverse of $n_{CPPR}$. The ejecta masses at $v_{eject} > u_{pH}$ (a) and $v_{eject} > \sqrt{2}u_{pH}$ (b) are shown as the percentage of the projectile mass. The dotted lines are linear functions obtained by the least squares method. The open symbols do not follow a linear trend and were not used in the fitting described above. The intersection point between the straight lines and the left Y-axis is the ejected mass at infinite spatial resolution.

Figure 15 is the same as Fig. 14b, except that the former shows the resolution effects on the initial position of the MM ejecta. Although we rejected particles initially located near the target surface, which are indicated in blue in the figure, to minimize the numerical artifacts resulting from artificial viscosity, it appears that position does not strongly depend on $n_{CPPR}$, indicating that the effect of shock smearing on ejection behavior of high-speed lightly shocked materials may not be as large as previously thought [DeCarli et al., 2007; DeCarli, 2013]. In fact, the mass of the MM ejecta, including the tracers initially located in the top five layers, is very close to the extrapolated results shown as the red dotted line in Fig. 16 at infinite $n_{CPPR}$ (Supplementary Materials S7). Figure 16 shows the ejected mass at >5 km/s as a function of the inverse of $n_{CPPR}$ for a vertical impact at 12 km/s. Although the spatial resolution in the SPH simulations is much lower than in the iSALE simulations, the total masses of the ejecta at >5 km/s calculated by the 3-D SPH model are similar to those of the 2-D iSALE model. This result implies that the SPH scheme, which is a Lagrangian code, is more suitable for solving for ejecta in the condensed phase.

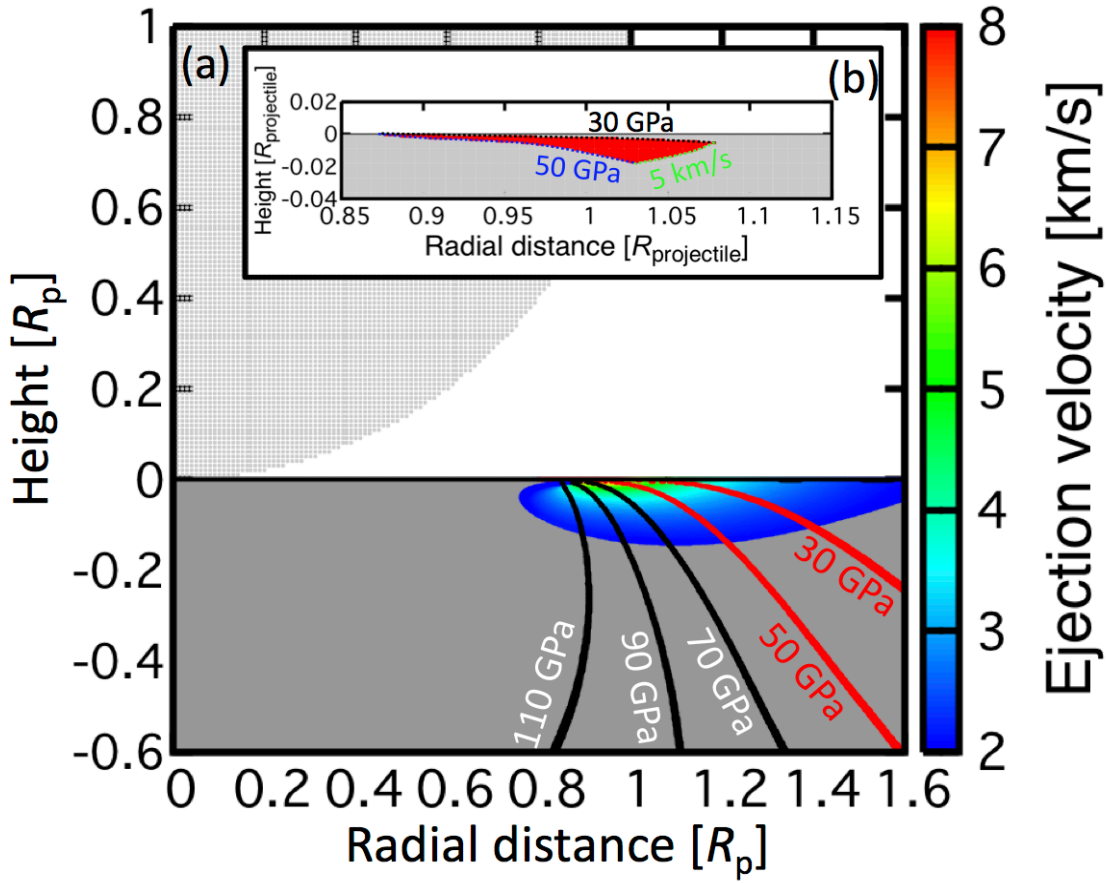

**Figure 14.** (a) Initial positions of the ejecta launched prior to $t = 1.4\ t_s$. The iSALE result with $n_{CPPR} = 1000$ for an impact at 12 km/s is shown. The color indicates the ejection velocity of each particle. The five isobaric lines for the peak pressures are also shown. (b) Initial position of the ejecta that meet the MM conditions (see Section 1). The three boundaries constrained by the range of peak pressure (30–50 GPa) and the escape velocity (>5 km/s) are indicated.

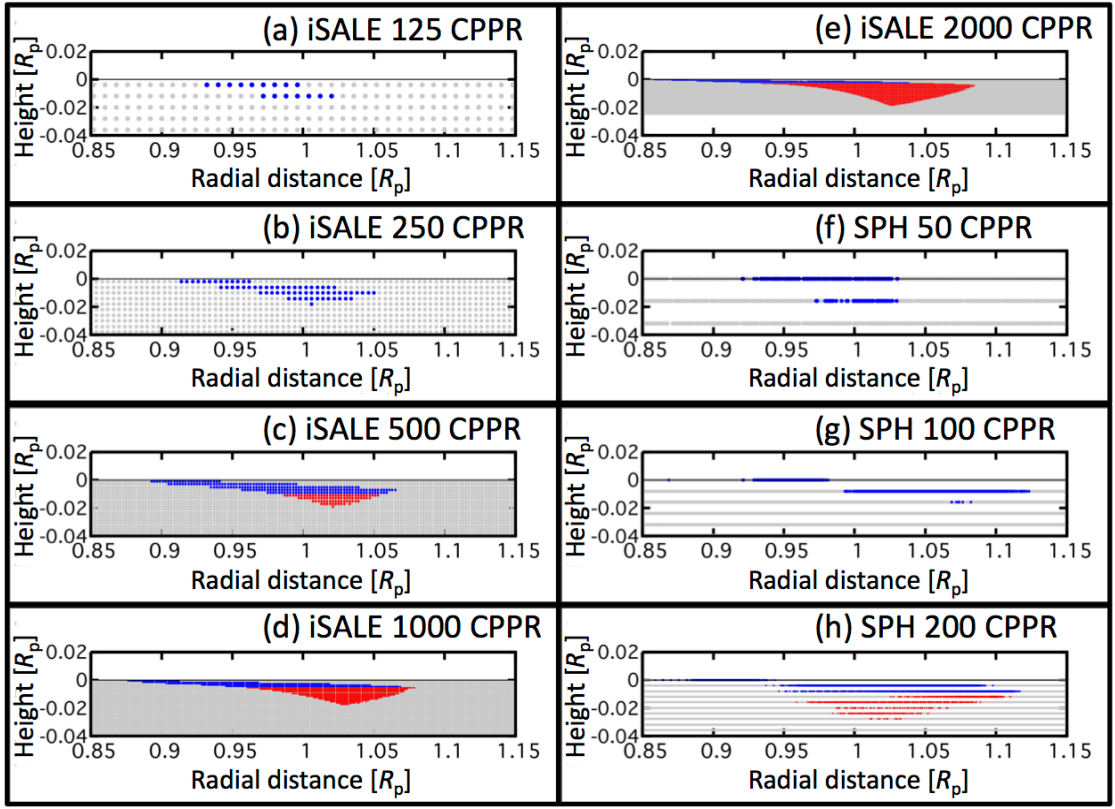

**Figure 15.** Same as Fig. 14b, but showing the effects of $n_{CPPR}$ and the numerical scheme on the launch position of the ejecta that meet the MM conditions. The iSALE results are shown in (a–e) and the SPH results are presented in (f–h). The $n_{CPPR}$ used in each calculation is indicated on the figure. Note that the particles initially located near the target surface are also shown as blue points, although these particles were not used in the analyses presented in this study.

However, the mass of MM ejecta strongly depends on $n_{CPPR}$. We cannot observe MM ejecta in the SPH simulations with $n_{CPPR} < 200$. In addition, the mass of MM ejecta in the SPH simulations with $n_{CPPR} = 200$ is 20 times smaller than that in the iSALE simulations with $n_{CPPR} = 1000$. In the case of the iSALE results, the masses of the MM ejecta with different $n_{CPPR}^{-1}$ are plotted as a linear function, allowing estimation of the mass at an infinite $n_{CPPR}$ by extrapolation. The difference between the results of our nominal model, $n_{CPPR} = 1000$, and the extrapolated value is within a factor of two. Thus, we largely presented the iSALE results with $n_{CPPR} = 1000$ throughout this paper.

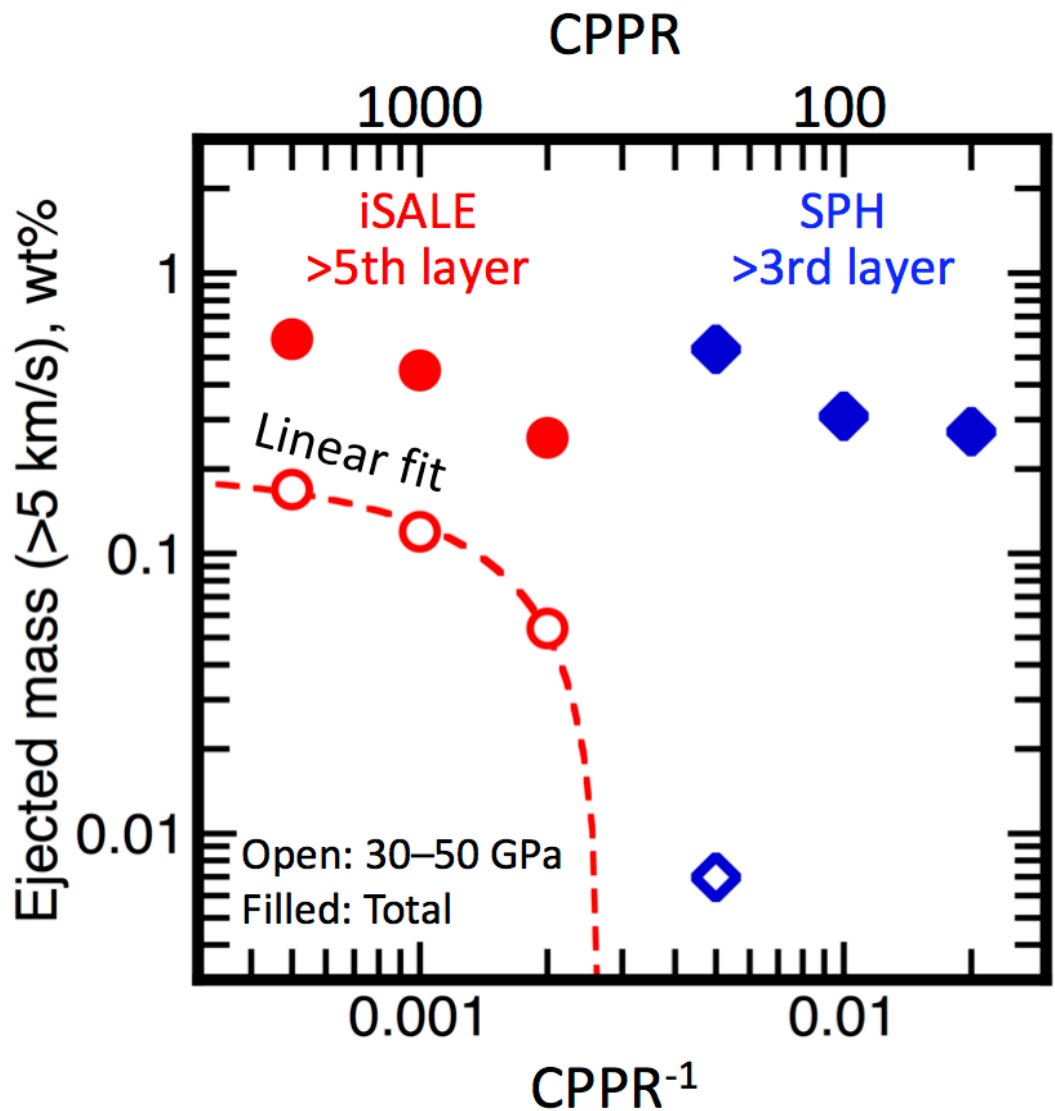

**Figure 16.** Mass of ejecta launched at >5 km/s as a function of the inverse of $n_{CPPR}$ on a log–log plot. The impact velocity is 12 km/s. The iSALE and SPH results are shown as red circles and blue squares, respectively. The filled and open symbols indicate the total mass and the mass that meets the MM conditions, respectively. The dotted red line is the linear function obtained by the least squares method, as in Fig. 14. Extrapolation of the straight line to the X-axis indicates that $n_{CPPR} > 370$ is required to resolve the ejecta that meet the MM conditions in the 2-D iSALE results.

Figure 17 shows the velocity effect on the ejected mass at $v_{eject}$ higher than the escape velocity of Mars. The total ejecta mass and mass of MM ejecta are shown in this figure. The mass of the MM ejecta is 0.01–0.1 wt% of the projectile mass at this velocity range. There is an optimum impact velocity to launch Martian meteorites of 9 km/s, whereby ~70 wt% of the ejected material at speeds of >5 km/s experienced $P_{peak}$ = 30–50 GPa. In contrast, only ~1 wt% of the whole ejecta is MM ejecta in the case of a 21 km/s impact. We also investigated the effects of impact velocity on the initial position of the MM ejecta (See Supplementary Materials S8).

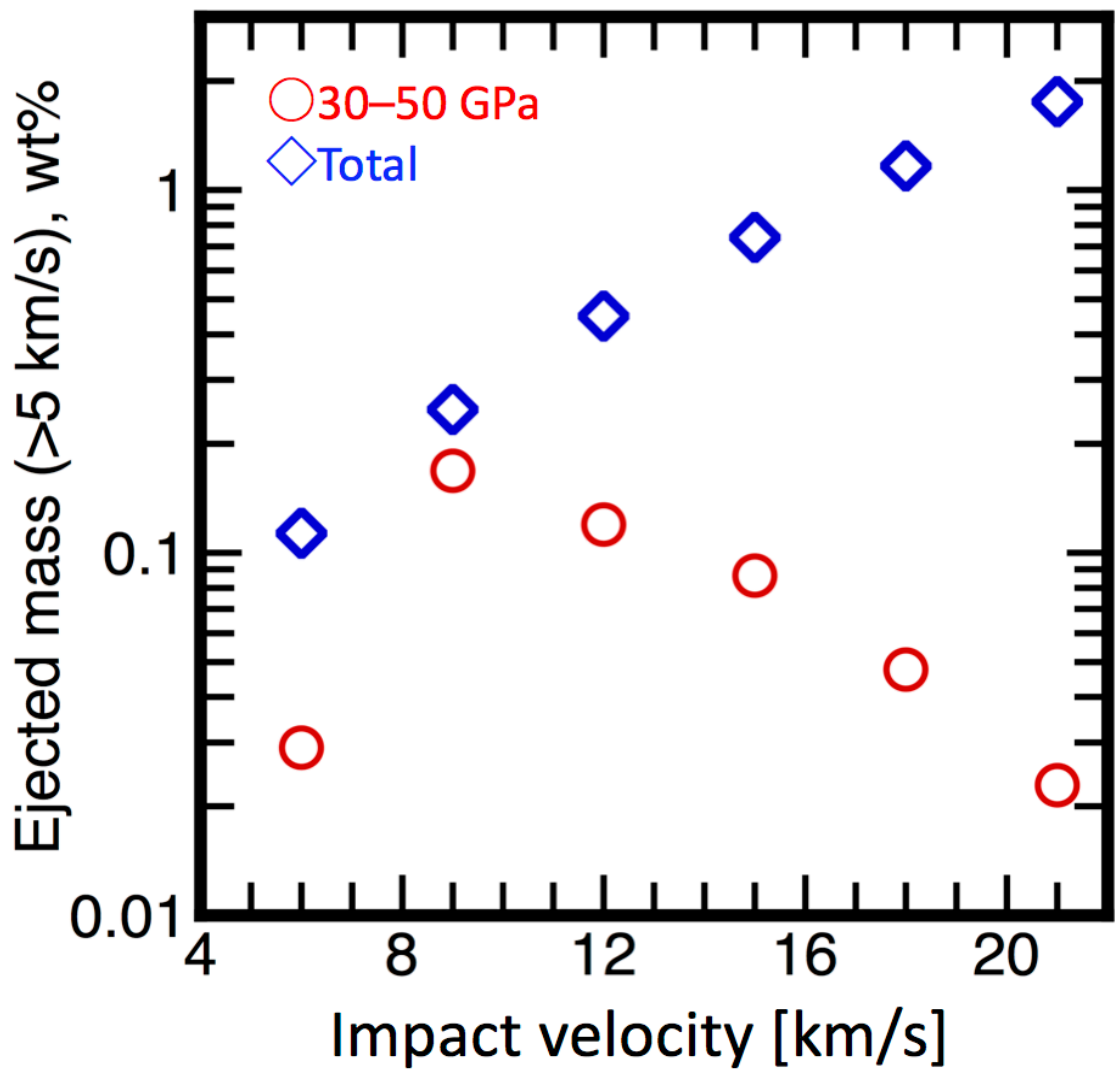

**Figure 17.** Masses of ejected materials launched at >5 km/s as a function of impact velocity. The total mass and the mass that meets the MM conditions are indicated as blue open squares and red open circles, respectively.

# 5. Discussion

## 5.1. Transition from jetting to spallation

In Section 4.4, we noted that the difference in angle between the incidence directions of the shock and expansion waves distinguishes spallation and normal excavation (Fig. 8c). In this section, we explore the difference between jetting and spallation processes. Figure 18a–d show schematic diagrams during blunt body penetration up to ~0.4 $t_s$ and the initiation of jetting and spallation. Figure 18a shows the case prior to the onset of jetting. The jetting initiates after shock detachment from the collision point, which is the location where the penetrating projectile and unshocked target converge. This is the time of origin of the propagation of the expansion wave, which can be calculated using standard jetting theory [e.g., Walsh et al., 1953; Kurosawa et al., 2015]. At the very early stages of pressure release, the expansion wave propagates inward in the radial direction shown in Fig. 18b. Thus, the travel direction of the expansion wave is almost opposite to that of the shock wave, resulting in an ejection velocity of ~$2u_{pH}$. The oblique convergence between the penetrating projectile and target causes a higher $P_{peak}$ and $u_{pH}$ than calculated by using the 1-D impedance match solution and Eq. (1), leading to an extremely high particle velocity of the jetted materials [e.g., Walsh et al., 1953; Kurosawa et al., 2015]. After the onset of jetting, the peak pressure gradually decays with increasing distance from the impact point. The expansion wave then mainly propagates downwards in the target. This downward-moving expansion wave produces the near-triangular shape of the released region as shown in Fig. 8a. The angle between the shock and expansion waves changes from ~180° to ~90° at the time shown in Figs 8b, 9, and 18c. Subsequently, the ejection flow driven by the shock release concentrates around ground level due to the difference in $u_{pH}$, which is a result of the decaying shock propagation, resulting in the late-stage acceleration of up to >$\sqrt{2}u_{pH}$ (Fig. 18d). The angle between the shock and expansion waves further changes from ~90° to ~0° at a later time, leading to normal excavation (Fig. 8c).

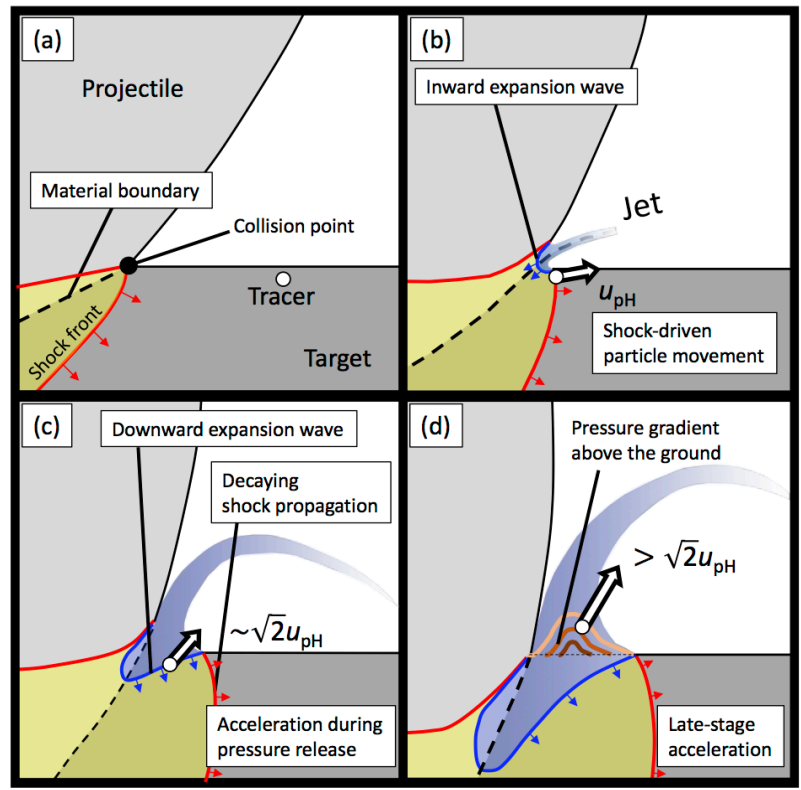

**Figure 18.** Schematic diagrams of the generation of jetted and spalled materials. These diagrams are drawn so that the location of the collision point, which can be calculated based on the kinematics of the penetrating projectile [e.g., Ang, 1990], is fixed even after the onset of jetting. The unshocked, shocked, and released materials are shown in grey, yellow, and blue, respectively. The released materials are defined as the materials where their temporal pressures fall below one-third of the peak pressures. (a) Prior to the onset of jetting. This situation is often referred to as the "regular regime" in jetting studies [e.g., Walsh, 1953]. (b) Immediately after the onset of jetting. An inward expansion wave is generated due to the shock detachment from the collision point. The materials incorporated into the ejection flow up to this time are expected to form ground-hugging hypervelocity jets. After the onset of jetting, a shock decay takes place. The tracer shown in (b) accelerates to $u_{pH}$ almost parallel to the target surface due to the weakened shock wave. (c) An expansion wave mainly propagates downward into the target. The tracer accelerates to $\sim\sqrt{2}u_{pH}$ due to isentropic release. (d) A pressure gradient in the root of the ejecta curtain is produced by piling up of the ejection flow (see the main text). The tracer further accelerates up to $>\sqrt{2}u_{pH}$.

*5.2. Effects of impact angle*

We only modeled vertical impacts in order to obtain a basic understanding of the impact-driven flow field within 1.5 $R_p$ from the impact point. The most likely impact angle in natural impact events is, however, 45° measured from the horizontal [e.g., Shoemaker, 1963]. Given that oblique impacts produce a 3-D hydrodynamic flow, we need to use a 3-D code with a similar spatial resolution to that employed in this study. Unfortunately, the use of iSALE-3-D [e.g., Elbeshausen et al., 2009] or other Eulerian codes to perform such calculations is prohibitively expensive. Although a higher spatial resolution than that used in this study is necessary to estimate the mass of the fast ejecta quantitatively, our 3-D SPH model may be a powerful tool for this objective in future studies. As such, we now discuss how the impact obliquity might qualitatively affect the ejection behavior.

Although oblique impacts lead to complex 3-D phenomena [e.g., Schultz and Gault, 1990], they can be broadly approximated as vertical impacts with translational motions parallel to the target surface. Thus, the velocity vector of the translational motion may be added to that of each ejecta particle obtained in this study. This leads to an increase in the ejecta mass that meets the MM conditions and an azimuthal anisotropy in the ejection velocity. This is qualitatively consistent with the results of Artemieva and Ivanov (2004), who showed that the mass of Martian meteorites produced during oblique impacts are one to two orders of magnitude higher than during vertical impacts, although the effects of shock smearing need be removed from their results. Consequently, the ejected mass at >5 km/s shown in Fig. 17 is a minimum estimate.

*5.3. Limitations of our models*

We only considered hydrodynamic motions during vertical impacts and neglected fragmentation processes in our modeling. The measured size of Martian meteorites and the shock duration are also important factors in terms of whether spallation can explain their launch [e.g., Melosh, 1984; Head et al., 2002; Artemieva and Ivanov, 2004; Baziotis et al. 2013]. The size of the ejected materials would depend on the peak pressure and strain rate during propagation of the stress wave [e.g., Melosh, 1984].

Recently, the Grady–Kipp fragmentation model [Johnson et al., 2016] or more advanced model [Melosh et al., 2017] were implemented into 2-D iSALE. Bowling et al. (2015) addressed the relation between the peak pressure and the dwell time distributions following an impact using 2-D iSALE. These studies may make it possible to address the size of the fast ejecta and to constrain the projectile size and the source crater.

We also neglected atmospheric effects in this study. Aerodynamic deceleration is expected to reduce the mass of Martian meteorites [e.g., Artemieva and Ivanov, 2004]. The magnitude of the speed reduction depends on the initial launch speed and angle, and the size of each fragment. The impact obliquity is also an important factor for estimating the speed reduction, because it significantly affects the launch speed and angle. If we could obtain a time-dependent stress field during oblique impacts and resultant launch speeds and angles using the 3-D SPH model at higher spatial resolution, the fragment size and aerodynamic interactions could be calculated computationally. However, such investigations are beyond the scope of this study.

### 5.4. Geological implications

The main finding of this study is that vertical impacts can accelerate shocked materials up to velocities of $>2u_{\text{pH}}$. This has a wide range of implications, not only for the acceleration of Martian meteorites, but also for material exchange amongst planetary bodies in satellite systems, including the Earth-Moon, Mars-satellites, Jovian, Saturnian, and Pluto–Charon systems [e.g., Melosh, 1984; Artemieva and Ivanov, 2004; Artemieva and Lunine, 2005; Stern, 2009; Chappaz et al., 2013; Ramsley and Head, 2013; Porter and Grundy, 2014]. In addition, our numerical models may have astrobiological implications, such as for the Panspermia [e.g., Melosh, 2003; Burchell et al., 2003; Price et al., 2013; Krijt et al. 2017; Lingam and Loeb, 2017].

Recently, JAXA (Japan Aerospace eXploration Agency) announced that the next target for a sample return mission is Phobos, one of the martian satellites. Phobos orbits Mars closer than any other moon and planet in our solar system. Given that high-speed ejecta (>4 km/s) from Mars could reach Phobos's orbit, Martian materials are likely to have accumulated on Phobos. Therefore, it would be possible to collect Martian rocks as well as Phobos' original rocks simultaneously in a single exploration mission, if the mixing ratio of Martian materials in Phobos's regolith is sufficiently high. An accurate estimate of the mixing ratio is necessary to ensure the correct specification of the onboard equipment for sample collection. Our findings should greatly improve the accuracy of these estimations, although the effects of impact obliquity, fragmentation, and aerodynamic interaction also need to be considered.

Another important implication of our work relates to the terrestrial origin of tektites. Tektites are natural glasses related to impact events [e.g., Koeberl, 1986]. Given that the ages of some tektites, including those from the Ivory Coast and Moldova, are similar to the ages of impact melts in terrestrial craters, it is thought that tektites have impact origins. Geochemical constraints on the origins of tektites have shown that they are produced from impact melts, and do not have any measureable projectile contamination [e.g., Koeberl, 1986]. In addition, they are likely to be launched at >6 km/s, because they traveled up to several hundred kilometers from the source craters to the collection sites [Wasson, 2015]. Jetting has been proposed as the mechanism for tektite production, based on an analytical model [Vickery, 1993]. However, Vickery (1993) rejected the jetting origin of the tektites because jetting should produce tektites that are a mixture of the projectile and target materials. In contrast, spalled materials from near the target surface only consist of target material. Our models show that fast materials traveling at ~10 km/s are also produced (Fig. 9b) during vertical impacts at 12 km/s, which is similar to the vertical component of the impact velocity onto Earth under typical impact conditions (17 km/s at 45° measured from the horizontal [e.g., Ito and Malhotra, 2006]). Given that the experienced shock pressures reach ~80 GPa, the materials would be melted after pressure release [e.g., Ahrens and O'Keefe, 1972]. Consequently, a spallation origin for tektites is consistent with both geochemical and physical constraints.

## 6. Conclusions

Impact spallation processes during vertical impacts were numerically modeled using the 2-D iSALE and 3-D SPH codes combined with the Tillotson EOS, which can approximate the hydrodynamic response of materials near the impact point. We have carefully assessed the effects of artificial viscosity on the ejection behavior through a series of hydrocode calculations over a wide range of spatial resolutions up to 2000 CPPR. This suggests that the uppermost three to five layers from the target surface lead to erroneously low peak pressures. We found that ejected materials can be accelerated to velocities higher than the ideal maximum for condensed materials determined from shock physics. Such high speeds are expected to be generated by a pressure gradient in the root of the ejecta curtain. The pressure gradient comes from the pileup of the ejection flow around the target surface due to the difference in the particle velocity produced by a decaying shock propagation. We refer to this new acceleration mechanism as "late-stage acceleration". Given that the results for vertical impacts apply to the production of lightly shocked high-speed ejecta, which we refer to as "MM ejecta", we have shown that: (1) the mass of MM ejecta is limited to 0.1–1 wt% of the projectile mass; (2) the initial depth of the MM ejecta is within ~2% of the projectile radius from the target surface; and (3) the impact velocity that is most favorable to the production of MM ejecta is 10 km/s. Although the effects of impact obliquity, fragmentation, and aerodynamic deceleration need to be addressed, our findings provide new insights into the processes of material exchange between planetary bodies.

**Acknowledgements**

We thank the developers of iSALE, including G. Collins, K. Wünnemann, B. Ivanov, J. Melosh, and D. Elbeshausen. The quick look of the iSALE results using the pySALEPlot tool written by Tom Davison greatly helped us to conduct the series of numerical simulations. We appreciate Hiroki Senshu and Shigeru Wakita for advice on the post-analysis using the particle tracking technique. We thank Koji Wada and Takafumi Matsui for stimulating this study. We also acknowledge useful discussions at a workshop on planetary impacts held at Kobe University. We thank Jay Melosh and an anonymous referee for their careful reviews that helped greatly improve the manuscript, and Francis Nimmo for helpful suggestions as an editor. The authors (KK, TO, and HG) are supported by JSPS KAKENHI Grant No. 17H02990. KK is supported by JSPS KAKENHI Grant Nos 17K18812, 15H01067, 26610184, and 25871212 and by the Astrobiology Center of the National Institutes of Natural Sciences, NINS (AB281026). HG is also supported by JSPS KAKENHI Grant No 15K13562 and by the NINS (AB28016).

**Supplementary Materials**

**Hydrocode modeling of the spallation process during hypervelocity impacts:**
**Implications for the ejection of Martian meteorites**

Kosuke Kurosawa[a,*], Takaya Okamoto[a], and Hidenori Genda[b]

[a]Planetary Exploration Research Center, Chiba Institute of Technology, 2-17-1, Tsudanuma, Narashino, Chiba 275-0016, Japan
[b]Earth–Life Science Institute, Tokyo Institute of Technology, 2-12-1 Ookayama, Meguro-ku, Tokyo 152-8550, Japan

*Corresponding author
Kosuke Kurosawa
Researcher, Planetary Exploration Research Center, Chiba Institute of Technology
Tel: +81-47-478-0320
E-mail: kosuke.kurosawa@perc.it-chiba.ac.jp

**S1. Irregular shock reflection region near the free surface**

Here, we discuss the "near-surface wave interaction" in the presence of a shock wave. The propagation speed of the rarefaction wave is the sum of the sound speed of the shocked materials and the parallel component of the particle velocity to the propagating direction of the rarefaction wave. This speed is faster than the propagation speed of the shock wave in general. Thus, the rarefaction wave is able to catch up with the outward-propagating shock wave, leading to a sudden decay of the shock wave. The time when the rarefaction wave catches up with the shock wave depends on the geometric configuration, especially the distance from the free surface. At the near-surface region, such wave interference occurs at a very early stage of the shock propagation.

The location of the boundary of the irregular shock reflection region can be calculated analytically, although point-source approximation is necessary. Figure S1a shows the calculated boundary plotted on the initial position of the tracer particles. The peak pressures are also shown in this figure. We assumed the burial depth to be 0.6 $R_p$ in this calculation, where $R_p$ is the radius of the impactor. The critical angle $\alpha^*$, which is the angle at the burial point between the horizontal and the irregular shock reflection boundary on the target surface (Fig. S1), is given by [e.g., Rosenbaum and Snay, 1956; Kamegai, 1986]:

$$\alpha^* = \tan^{-1}\left[\frac{\sqrt{C_R^2 - (V_s - u_p)^2}}{V_s}\right], \text{(S1)}$$

where $C_R$, $V_s$, and $u_p$ are the sound speed of the shocked materials, shock speed, and particle velocities, respectively. Figure S2 shows $\alpha^*$ as a function of peak pressure calculated using the Tillotson EOS for granite, showing that $\alpha^*$ can be approximated to be 35° over a wide range of peak pressures. This result is consistent with the study of Melosh et al. (2017). Thus, we used 35° as the $\alpha^*$ value to calculate the boundary.

From simple geometric considerations, the distance $R$ from the burial point to the irregular shock reflection boundary at any angle $\alpha$ ($< \alpha^*$) is expressed by the following differential equation [Kamegai, 1986]:

$$R\frac{d\alpha}{dR} = -\tan\phi. \text{(S2)}$$

The angle $\phi$ corresponds to the angle between the shock and rarefaction waves. Importantly, $\phi$ is different from the angle between the shock and expansion waves. If the relationship between peak pressures $P_{peak}$ are known, effects of near-surface wave interaction are ignored, and the distance from the impact point $r$ is known, we can integrate Eq. S2 and obtain the position of the boundary (Fig. S1).

Figure S3 shows the $P_{peak}$–$r$ relationships for three different initial regions obtained by the iSALE results with $n_{CPPR}$ = 500 at 12 km/s. These three regions are in the down region from the impact point at 5°±3°, 45°±3°, and 75°±3°. The 45° and 75° directions have similar trends in $P_{peak}$–$r$. Therefore, the 75° direction is completely unaffected by the near-surface wave interaction. By using the $P_{peak}$–$r$ relationship for the 75° direction, the irregular shock reflection boundary (red line in Fig. S1) was calculated through integration of Eq. S2. The $P_{peak}$–$r$ relationship for the 5° direction differs significantly from the other regions. $P_{peak}$ suddenly decreases at ~0.8 $R_p$ with increasing $r$. Strictly speaking, we cannot define a certain value of the burial depth because the point-source approximation is not valid near the impact point. Nevertheless, a burial depth of 0.6 $R_p$ may be a good approximation as shown in Fig. S3.

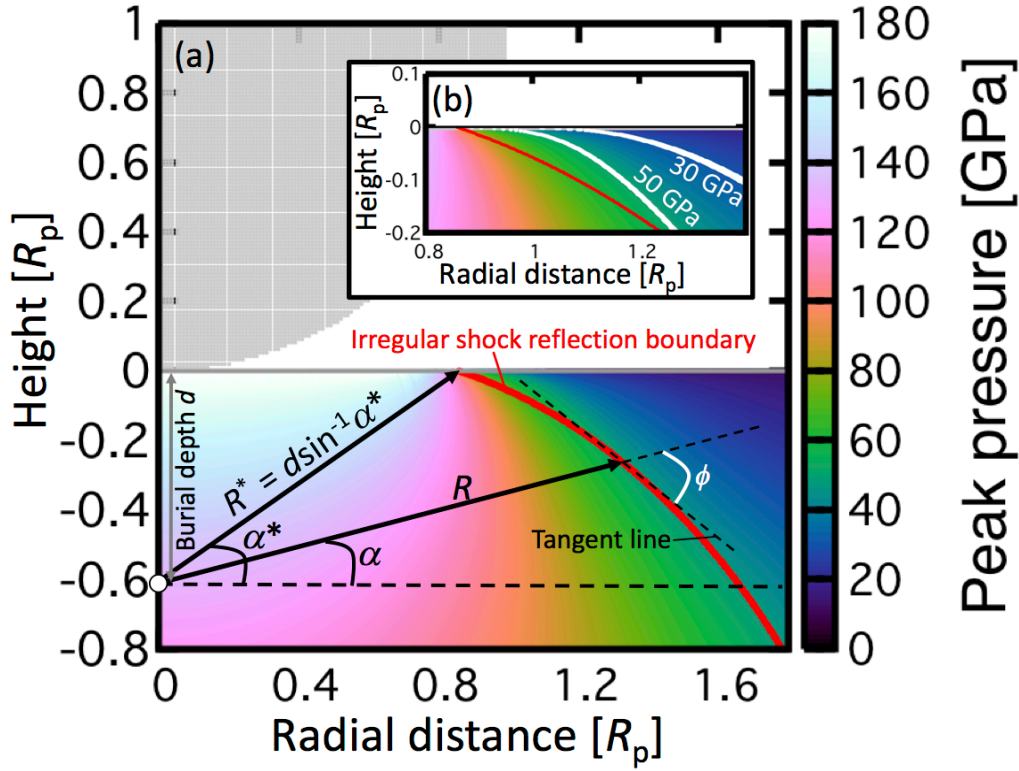

**Figure S1.** (a) The location of the irregular shock reflection boundary (thick red line) on a provenance plot of the tracers highlighted in color depending on peak pressures. Graphical definitions of the variables in Eqs S1 and S2 are shown. (b) Same as panel (a), except that two isobaric lines for $P_{peak}$ = 30 and 50 GPa (white lines) are shown.

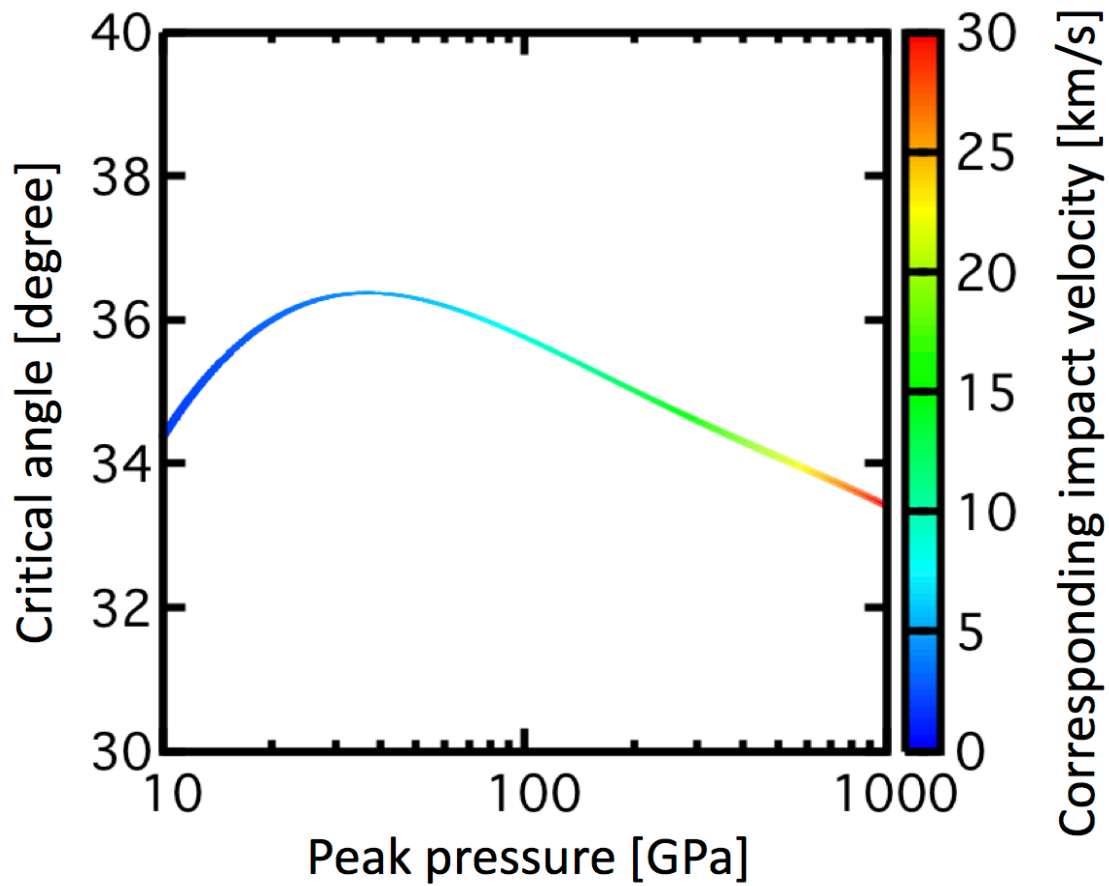

**Figure S2.** The critical angle $\alpha^*$ as a function of peak pressure. The colors indicate the corresponding impact velocity calculated to be twice of the particle velocity immediately after the shock wave passage.

Figure S2b is the same as Fig. S2a, except that two isobaric lines for $P_{peak}$ = 30 and 50 GPa are shown, along with the irregular shock reflection boundary, highlighting that the entire ejecta that experienced $P_{peak}$ of 30–50 GPa comes from the irregular shock reflection region. It should be noted that the rise time of the pressure is much shorter than the characteristic time of the projectile penetration $t_s$ (Fig. 12), despite the rarefaction wave having already caught up with the shock wave.

The shape of the irregular shock reflection boundary *appears* to produce a strong pressure gradient directed upward and outward, because the materials inside the boundary experience much higher peak pressures than those within the near-surface region (Fig. S3). However, this is not correct. We show that the actual direction of the pressure gradient after the shock wave passage in the near-surface region is mostly upward and inward as shown in Figs 8 and 9. The pressure gradient directed upward and outward is only produced at the root of the ejecta curtain and only by the late-stage acceleration, which results from the material pileup in the ejection flow after the shock-release sequence (see Section 4.5 in the main text).

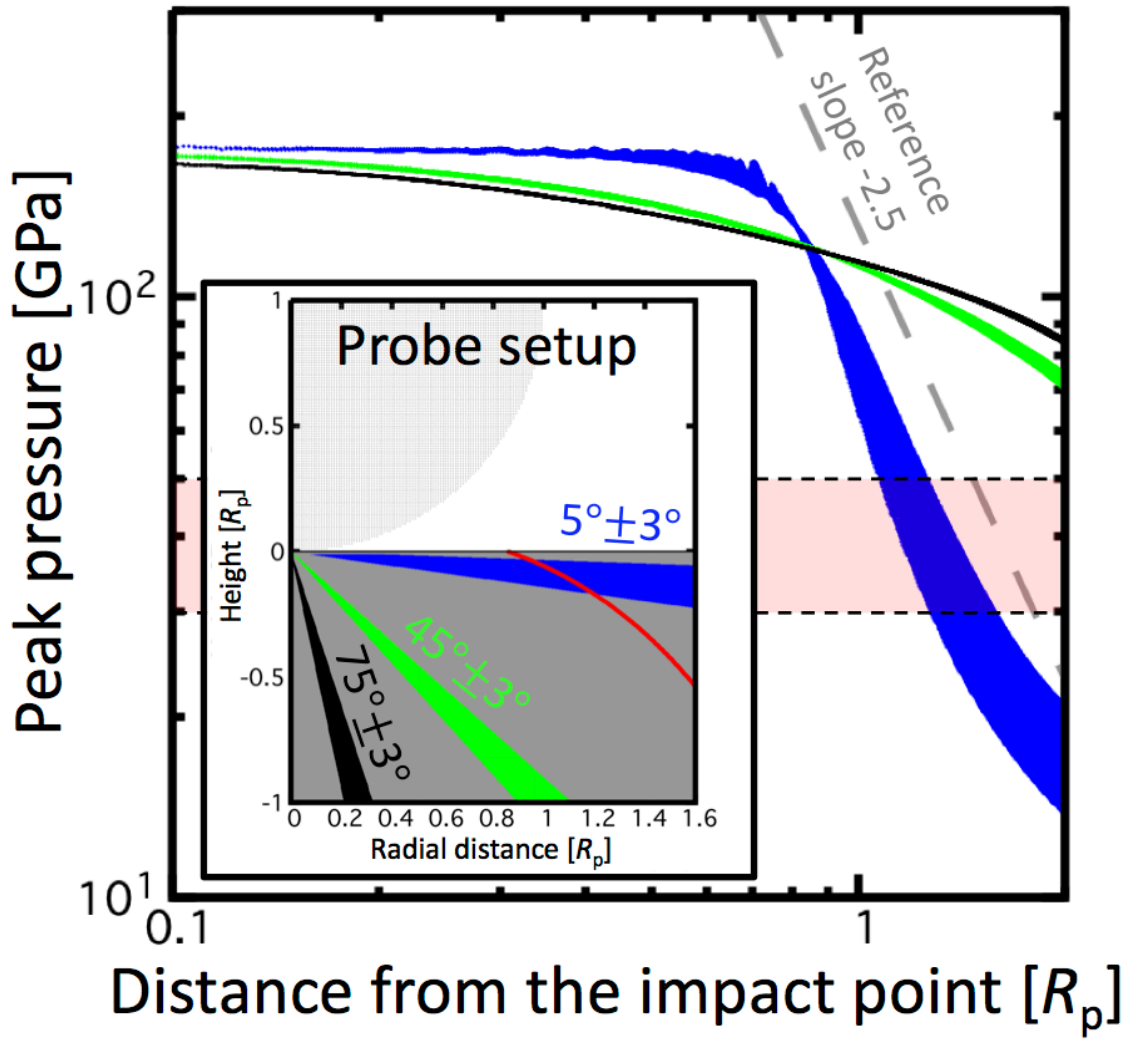

**Figure S3.** The peak pressure distribution as a function of the distance from the impact point. To produce the plot, we used three pressure probes directed in different directions as shown in the inset. The red line in the inset is the irregular shock reflection boundary, which is the same as in Fig. S1. The red shaded region corresponds to a range of peak pressure from 30 to 50 GPa. The peak pressure distribution with respect to the distance from the impact point is often expressed as a power law. The grey dotted line is a reference slope with an exponent of –2.5.

**S2. The contribution of the thermal component to the peak pressure**

For the Tillotson EOS, the pressure $P$ is given by a function of both density $\rho$ and internal energy $E$. In the compressed region ($\rho/\rho_0 > 1$, where $\rho_0$ is the density at the reference state), the pressure is expressed by the sum of the thermal ($P_{\text{thermal}}$) and cold components ($P_{\text{cold}}$) as follows [Tillotson, 1962]:

$$P = P_{\text{thermal}}(E, \rho) + P_{\text{cold}}(\rho), \qquad (S3)$$

where

$$P_{\text{thermal}}(E, \rho) = \left[ a + \frac{b}{\left( \frac{E}{E_0 \eta^2} + 1 \right)} \right] \rho E, \qquad (S4)$$

$$P_{\text{cold}}(\rho) = A\mu + B\mu^2, \qquad (S5)$$

where $\eta = \frac{\rho}{\rho_0}$, $\mu = \eta - 1$, and $a$, $b$, $A$, $B$, and $E_0$ are the Tillotson parameters listed in Table 1. Figure S4 shows $P_{\text{peak}}$ as a function of the particle velocity behind the shock wave $u_{\text{pH}}$ and contributions from $P_{\text{thermal}}$ and $P_{\text{cold}}$. At the range of $P_{\text{peak}}$ in this study (red shaded region), the contribution of $P_{\text{thermal}}$ is dominant. The particle velocity behind the shock $u_{\text{pH}}$ reaches >2 km/s, which cannot be neglected when we consider material ejection from the irregular shock reflection region.

Pressure release from the peak shock state initiates immediately after the shock arrival, because the rarefaction wave has already caught up with the shock wave. The pressure release can be approximated as an adiabatic process because of the high pressure. According to thermodynamic principles, the change in the internal energy d$E$ during adiabatic release can be expressed by:

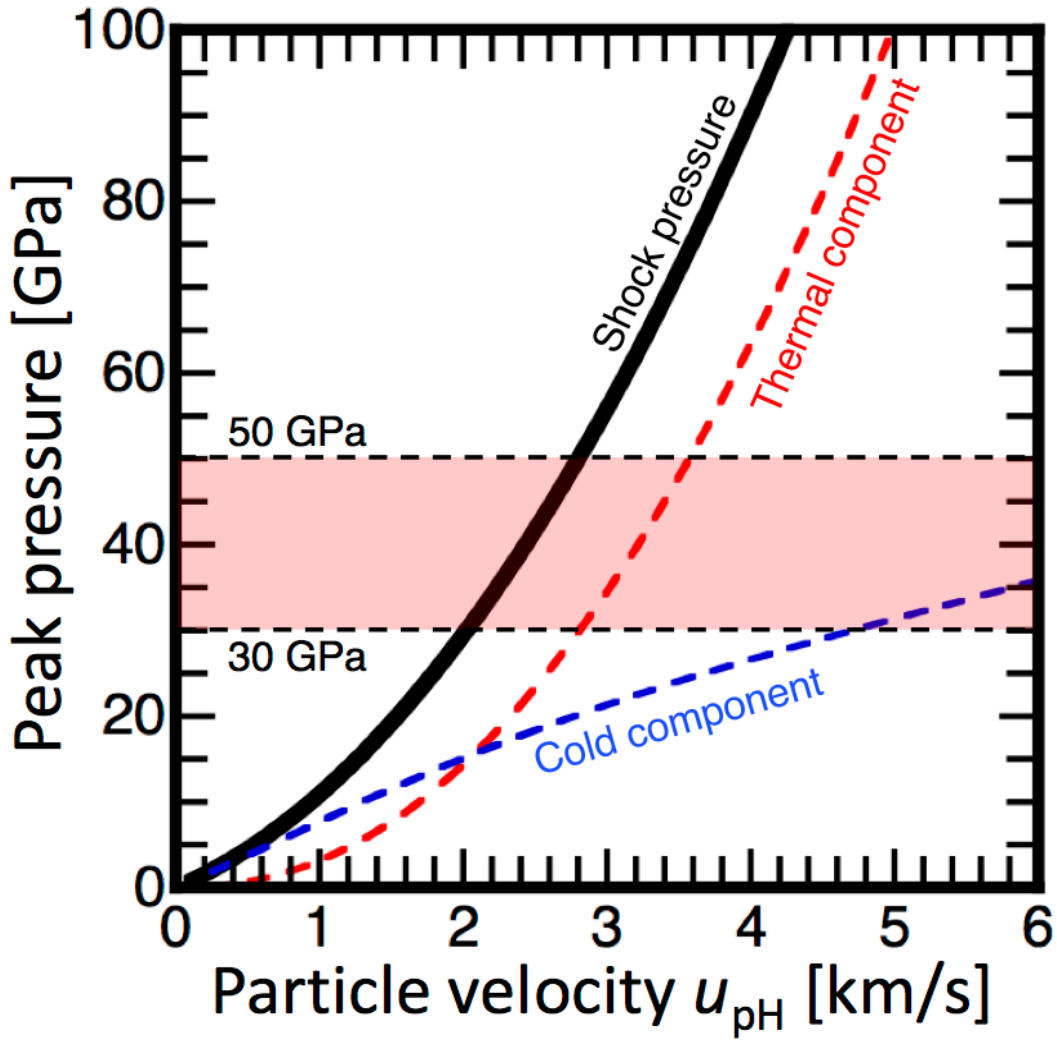

**Figure S4.** The peak pressure as a function of the particle velocity immediately after the shock wave passage as calculated by the Tillotson EOS. The red and blue dotted lines are the thermal and cold component defined by Eqs S4 and S5, respectively. The red shaded region corresponds to the range of peak pressure from 30 to 50 GPa as in Fig. S3.

$$\mathrm{d}E = \frac{P}{\rho^2}\mathrm{d}\rho, \qquad\qquad (S6)$$

where $P$ and $\rho$ are the temporal pressure and density during the release, respectively. Thus, the occurrence of fast adiabatic expansion to the free surface would be expected to relax the highly compressed state because adiabatic expansion effectively reduces the thermal pressure as predicted by Eqs S4 and S6.

## S3. The effects of the material difference on the 2-D iSALE results

We conducted another iSALE run with the Tillotson parameters for basalt [Benz and Asphaug, 1999] to investigate the effects of material differences on ejection behavior. We assumed that a basalt impactor hits a flat basalt target. The cumulative mass–velocity relationship from the iSALE results for a vertical impact at 12 km/s was derived to compare with the results for granite. Figure S5 shows the cumulative mass of the ejected materials at a given velocity as a function of the ejection velocity. The materials that experienced high $P_{\text{peak}}$, in the case of basalt, have somewhat higher ejection velocities than in the case of granite. Nevertheless, the material difference does not strongly affect the mass–velocity relationship at 30–50 GPa. This result allows us to apply the results obtained in this study to lightly shocked, high-speed basaltic ejecta.

## S4. The accuracy of the tracer tracking in the grid-based code

The tracer motion in iSALE is calculated using linear interpolation of the surrounding nodal velocities. This often causes "tracer drift" into neighboring materials [Davison et al., 2016], possibly leading to a deviating motion from the flow velocity in each computational cell. Figure S6 shows a comparison between the tracer motions and the raw grid data. This result clearly shows that the tracer motions obtained in our model closely follow the raw grid data.

## S5. The reliable depth of SPH simulations

Figure S7 is the same as Fig. 6 in the main text, except that the initial depth is 2.0%–2.5% of $R_{\text{p}}$ from the target surface and only the SPH results are shown. This depth corresponds to the 4[th] and 7[th] layers for SPH with $n_{\text{CPPR}} = 100$ and 200, respectively, where $n_{\text{CPPR}}$ is the number of cells per projectile radius. Although SPH particles with $n_{\text{CPPR}} = 200$ experience a wider range of $P_{\text{peak}}$ than those with $n_{\text{CPPR}} = 100$, the results are in good agreement. This figure indicates that SPH particles initially located below the 3rd layer are not affected by severe shock smearing.

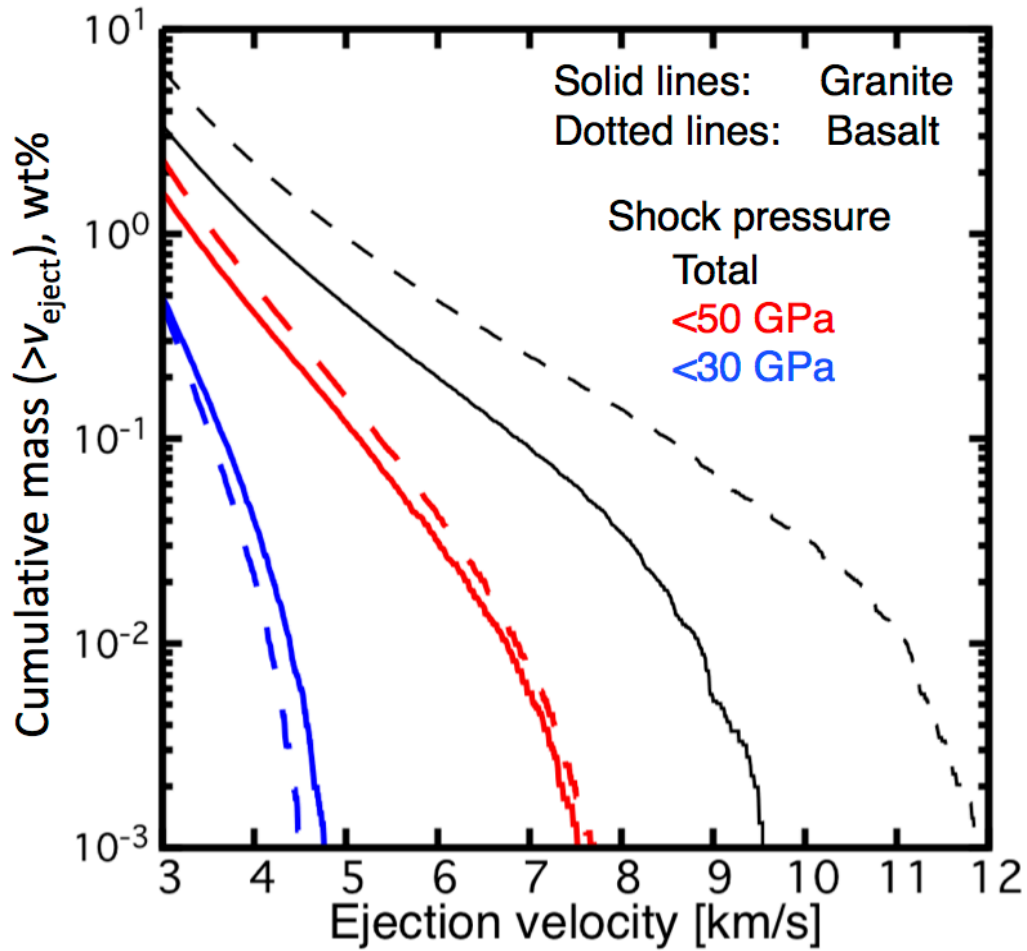

**Figure S5.** The cumulative mass of ejecta at a given ejection velocity as a function of the ejection velocity. The solid and dotted lines are the results for granite and basalt, respectively. The ejecta are classified into three groups (blue, red, and black) depending on the peak pressure experienced as shown in the figure.

## S6. Relationship between the ejection velocities and the peak pressure at a depth beneath the irregular reflection region

Here, we test the Melosh spallation model [Melosh, 1984, 1985a] using the iSALE results. As discussed in Section 2 in the main text, the spallation model cannot be applied to the Martian meteorites launch in principle because the key assumption, which is that a propagating stress pulse are described as a triangular wave with a finite rise time, is not valid in the case of the presence of shock waves. Nevertheless, we conducted an additional analysis by using the previous model in this section because the spallation model has been widely accepted. Figure S8b shows the ejection velocities as a function of the peak pressures without the surface effects $P_{free}$. We derived $P_{free}$ as a function of the distance from the impact point using the peak pressure well below the irregular shock reflection region (black line on Fig. S3 in Supplementary Materials). For comparison, the ejection velocities $v_{ej}$ against the actual peak pressures $P_{max}$ is also shown in the Fig. S8a. The iSALE result with $n_{CPPR} = 2000$ for an impact at 12 km/s are only shown. If the spallation model is applicable to the case of the presence of shock waves, the line pertaining to $v_{ej} = 2u_{pH}$ is expected to become the limiting velocity in Fig. S8b. Our test clearly shows that a small part of tracer particles is plotted above the line, suggesting that the spallation model is only applicable far from the impact point as mentioned in Melosh (1984).

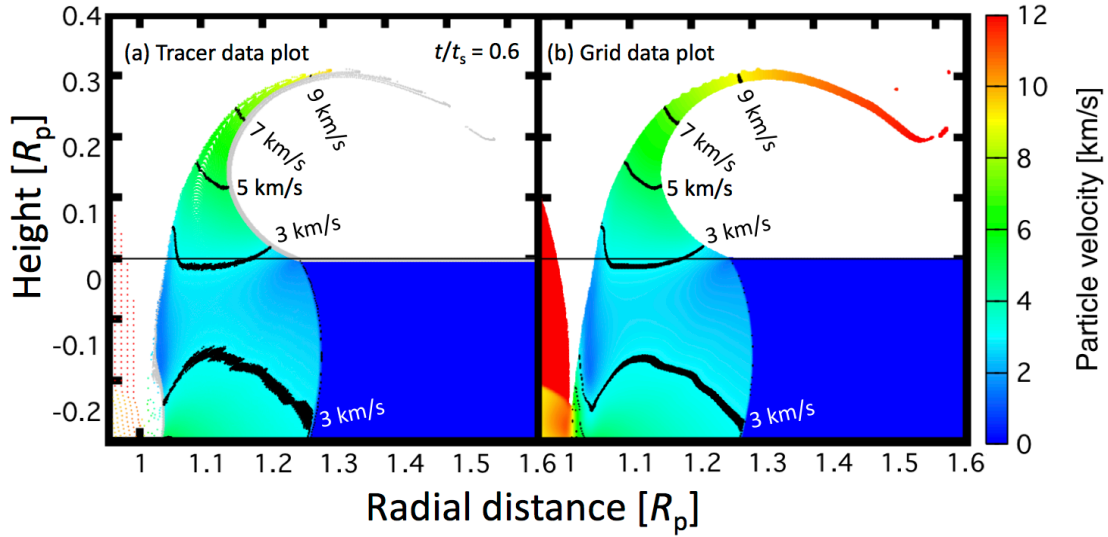

**Figure S6.** Snapshots of the iSALE results with $n_{CPPR} = 1000$ for a vertical impact at 12 km/s. The time $t$ is indicated in the figure, normalized to the characteristic time for projectile penetration $t_s = D_p/v_{imp}$, where $D_p$ and $v_{imp}$ are the projectile diameter and impact velocity, respectively. (a) The tracers are colored depending on the plotted particle velocities. (b) The grid data are plotted. The flow velocity in each computational cell is colored with the same color used in panel (a). Velocity contours are also shown in panels (a) and (b).

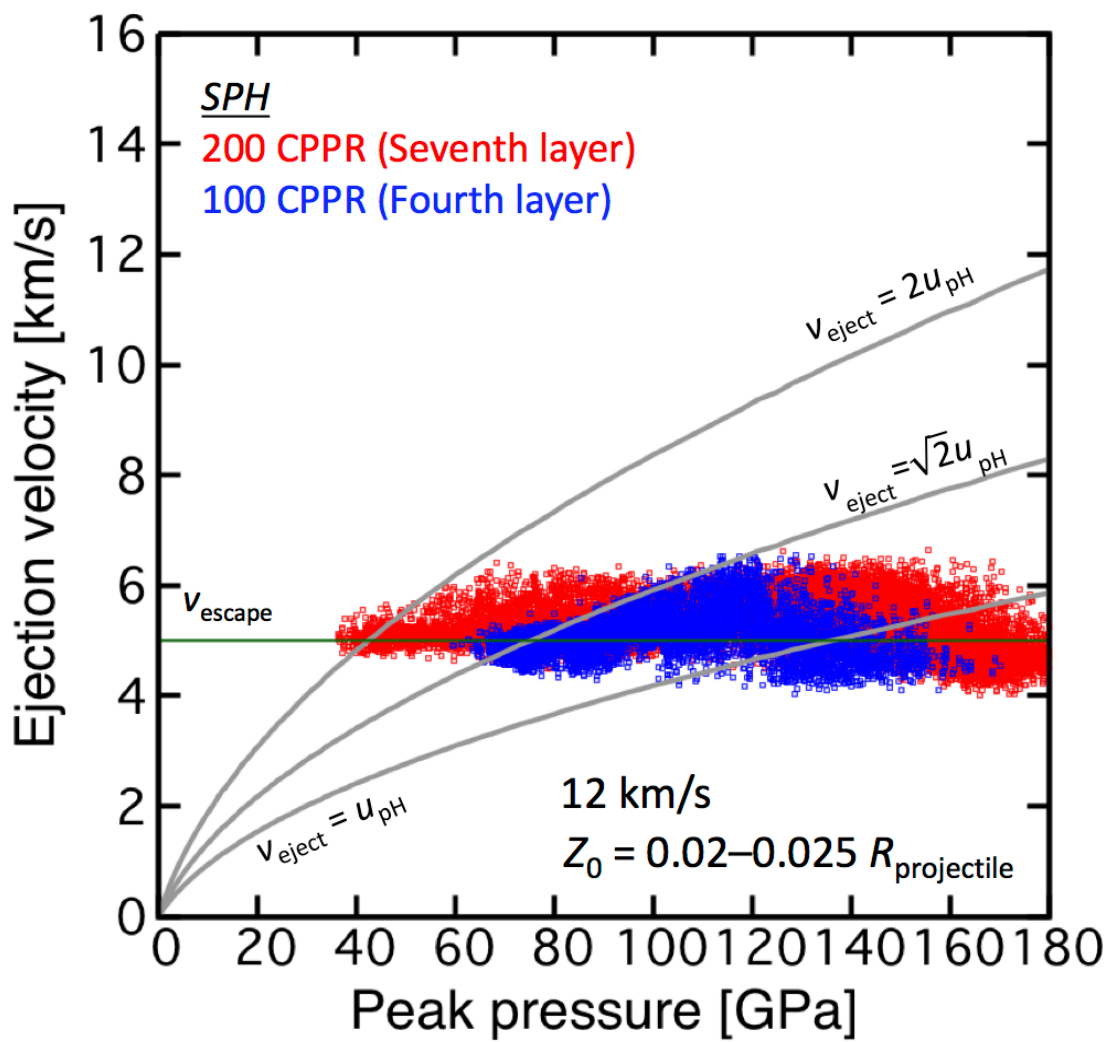

**Figure S7.** Ejection velocities as a function of peak pressures, but only showing the particles initially located at depths of 2.0%–2.5% of the projectile radius. The 3-D SPH results with $n_{CPPR}$ = 100 and 200 are only plotted.

## S7. Resolution dependence on the mass of Martian meteorite (MM) ejecta

We rejected the tracers initially located in the top five layers to obtain the mass of ejecta that meets the MM ejecta criteria, as described in Section 4.2 in the main text. The contribution of such unreliable tracers would reduce with increasing $n_{CPPR}$. Figure S9 is the same as Fig. 16 in the main text, except that the former also shows the mass of ejecta that meet the conditions with peak pressures $P_{peak} = 30$–$50$ GPa and ejection velocities $v_{eject} > 5$ km/s (hereafter, referred to as MM ejecta in Supplementary Materials), including the tracers initially located within the top five layers (red triangles) along with the results presented in the main text (blue squares). The masses of MM ejecta with different $n_{CPPR}^{-1}$ plot as a linear function at $n_{CPPR} > 500$, regardless of whether the top five layers are rejected or not. The extrapolation of the linear function to the Y-axis provides the mass of MM ejecta at infinite $n_{CPPR}$. Again, the extrapolated results with or without the rejection are in good agreement.

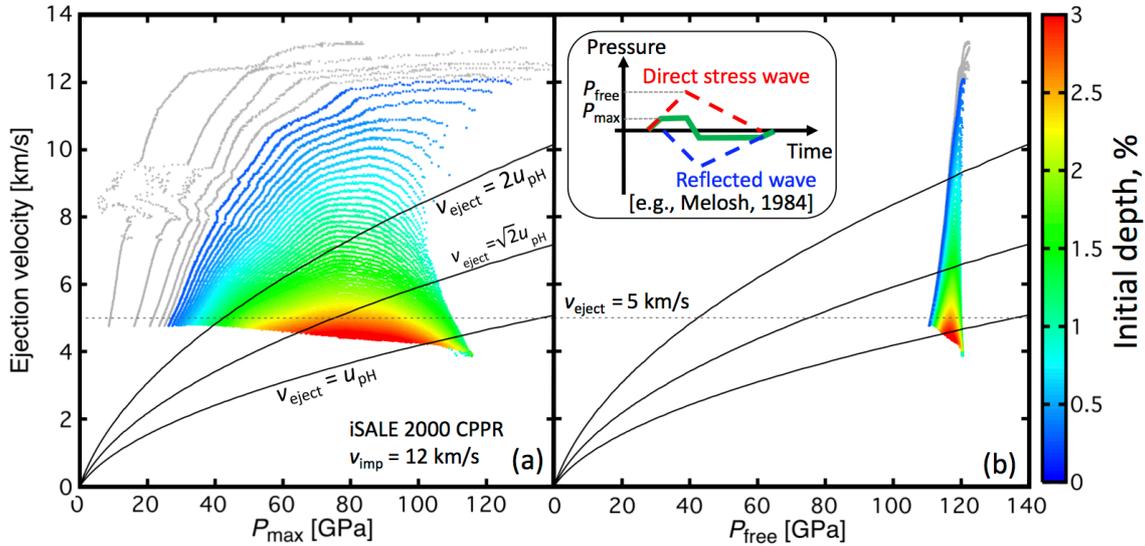

**Figure S8.** Ejection velocities as a function of $P_{max}$ (a) and $P_{free}$ (b). The colors indicate the initial depth expressed as a percentage of the projectile radius. The escape velocity of Mars $v_{escape}$ and the curves for $v_{eject} = u_{pH}$, $\sqrt{2}u_{pH}$, and $2u_{pH}$ are also shown as guides (see Section 2). The inset shows a schematic illustration of the pulse shape in the near-surface wave interaction proposed by Melosh (1984).

## S8. Velocity dependence on the initial location of Martian meteorite (MM) ejecta

Figure S10 shows the effects of impact velocity on the initial position of the MM ejecta. The initial position gradually moves outward in a radial direction with increasing impact velocity, because a higher impact velocity leads to a higher $P_{peak}$ at a given position. The initial depth is shallower than $0.02R_p$ from the target surface, irrespective of impact velocity.

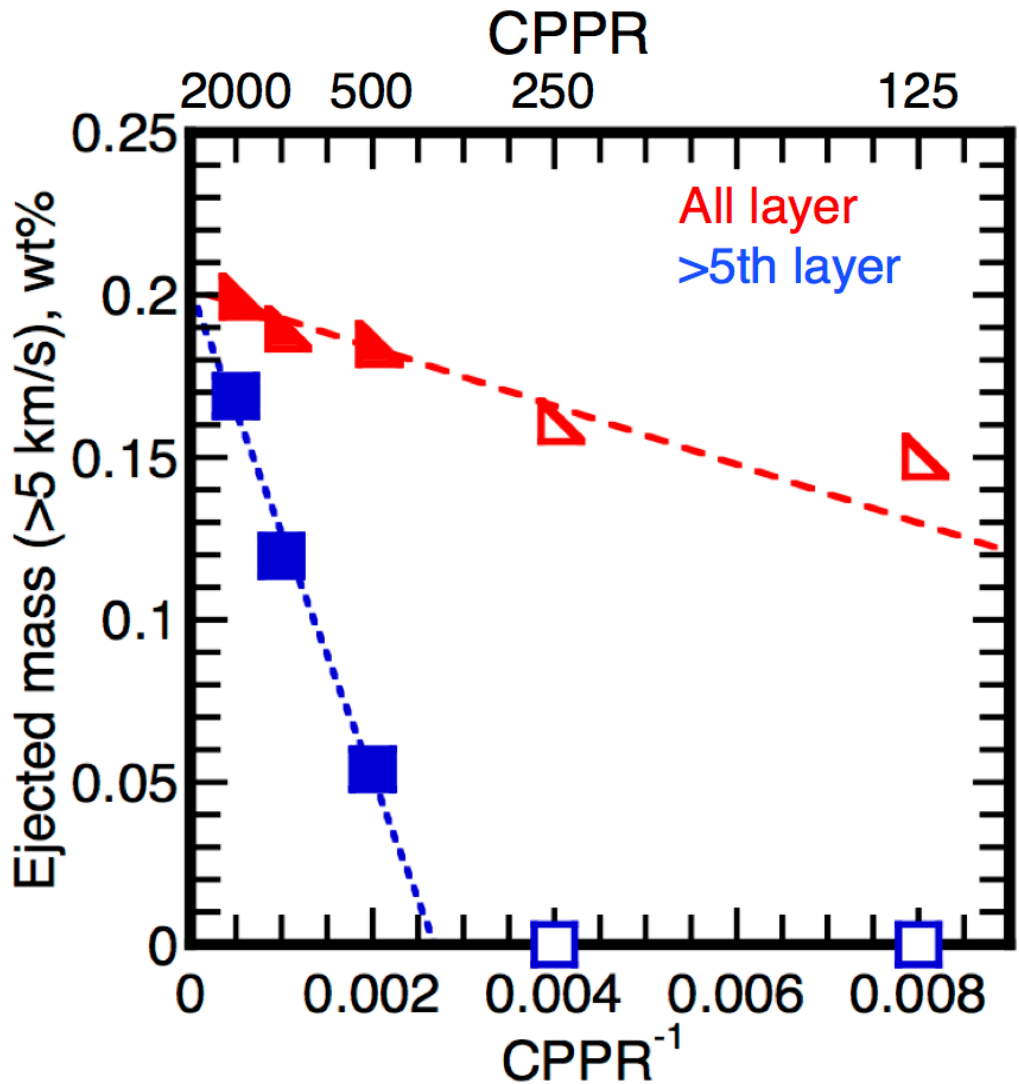

**Figure S9.** Mass of ejecta launched at >5 km/s as a function of the inverse of $n_{CPPR}$ on a linear–linear plot. The impact velocity is 12 km/s. The iSALE results are only shown. The blue squares and red triangles are the mass with or without tracer rejection, respectively. The dotted lines are linear functions obtained using the least-squares regression method. The filled symbols were used in this regression.

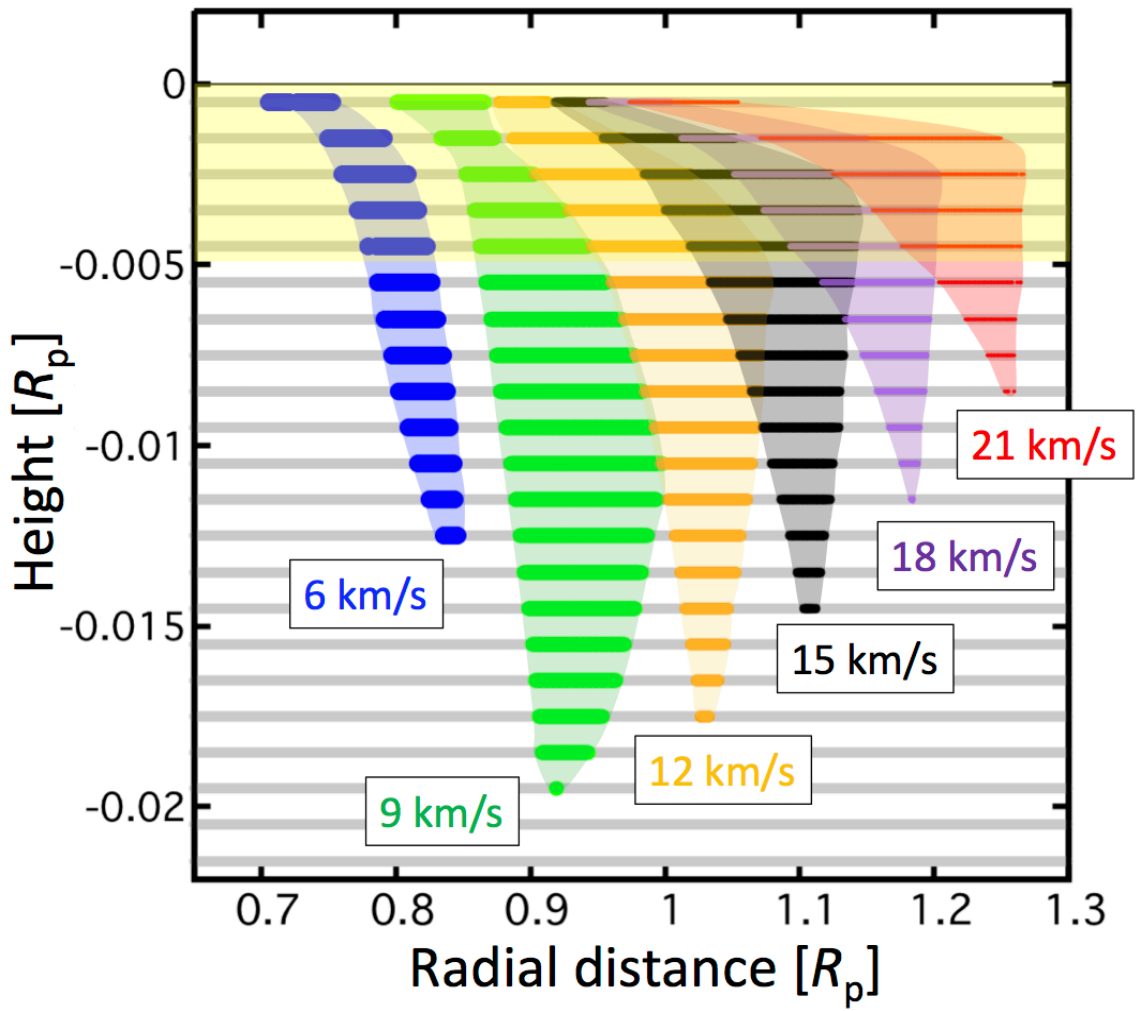

**Figure S10.** The initial position of the MM ejecta. The impact velocities used in the calculations are indicated on the figure. The iSALE results with $n_{CPPR} = 1000$ are presented. The yellow shaded region indicates the top five layers, which were not used in the ejecta mass estimate. Note that the vertical scale is exaggerated.